\RequirePackage{lineno}

\documentclass[aps,prd,preprint,superscriptaddress,showpacs,floatfix,nofootinbib]{revtex4}
\linenumbers
\linenumberdisplaymath
\usepackage{comment}
\usepackage{graphicx}
\usepackage{bm}
\usepackage{multirow}

\newcommand{\MET}{ $/\!\!\!\!\!E_{T}$ }
\newcommand{\METsig}{ $S_{/\!\!\!\!\!E_{T}}$ }
\newcommand{\METtext}{/\!\!\!\!\!E_{T}}
\newcommand{\METraw}{E_{T}^{\rm{raw}}\!\!\!\!\!\!\!\!\!\! / \;\;\;\;}
\newcommand{\METvertex}{E_{T}^{\rm{vertex}}\!\!\!\!\!\!\!\!\!\!\!\!\!\!\! / \;\;\;\;\;\;\;\;}

\begin{document}

\preprint{CDF/PUB/EXOTIC/CDFR/****}
\preprint{PRD draft version 4.3}

\title{Search for Standard Model Higgs Boson Production in Association
 with a $W$ Boson at CDF}

\affiliation{Institute of Physics, Academia Sinica, Taipei, Taiwan 11529, Republic of China} 
\affiliation{Argonne National Laboratory, Argonne, Illinois 60439, USA} 
\affiliation{University of Athens, 157 71 Athens, Greece} 
\affiliation{Institut de Fisica d'Altes Energies, ICREA, Universitat Autonoma de Barcelona, E-08193, Bellaterra (Barcelona), Spain} 
\affiliation{Baylor University, Waco, Texas 76798, USA} 
\affiliation{Istituto Nazionale di Fisica Nucleare Bologna, $^{aa}$University of Bologna, I-40127 Bologna, Italy} 
\affiliation{University of California, Davis, Davis, California 95616, USA} 
\affiliation{University of California, Los Angeles, Los Angeles, California 90024, USA} 
\affiliation{Instituto de Fisica de Cantabria, CSIC-University of Cantabria, 39005 Santander, Spain} 
\affiliation{Carnegie Mellon University, Pittsburgh, Pennsylvania 15213, USA} 
\affiliation{Enrico Fermi Institute, University of Chicago, Chicago, Illinois 60637, USA}
\affiliation{Comenius University, 842 48 Bratislava, Slovakia; Institute of Experimental Physics, 040 01 Kosice, Slovakia} 
\affiliation{Joint Institute for Nuclear Research, RU-141980 Dubna, Russia} 
\affiliation{Duke University, Durham, North Carolina 27708, USA} 
\affiliation{Fermi National Accelerator Laboratory, Batavia, Illinois 60510, USA} 
\affiliation{University of Florida, Gainesville, Florida 32611, USA} 
\affiliation{Laboratori Nazionali di Frascati, Istituto Nazionale di Fisica Nucleare, I-00044 Frascati, Italy} 
\affiliation{University of Geneva, CH-1211 Geneva 4, Switzerland} 
\affiliation{Glasgow University, Glasgow G12 8QQ, United Kingdom} 
\affiliation{Harvard University, Cambridge, Massachusetts 02138, USA} 
\affiliation{Division of High Energy Physics, Department of Physics, University of Helsinki and Helsinki Institute of Physics, FIN-00014, Helsinki, Finland} 
\affiliation{University of Illinois, Urbana, Illinois 61801, USA} 
\affiliation{The Johns Hopkins University, Baltimore, Maryland 21218, USA} 
\affiliation{Institut f\"{u}r Experimentelle Kernphysik, Karlsruhe Institute of Technology, D-76131 Karlsruhe, Germany} 
\affiliation{Center for High Energy Physics: Kyungpook National University, Daegu 702-701, Korea; Seoul National University, Seoul 151-742, Korea; Sungkyunkwan University, Suwon 440-746, Korea; Korea Institute of Science and Technology Information, Daejeon 305-806, Korea; Chonnam National University, Gwangju 500-757, Korea; Chonbuk National University, Jeonju 561-756, Korea} 
\affiliation{Ernest Orlando Lawrence Berkeley National Laboratory, Berkeley, California 94720, USA} 
\affiliation{University of Liverpool, Liverpool L69 7ZE, United Kingdom} 
\affiliation{University College London, London WC1E 6BT, United Kingdom} 
\affiliation{Centro de Investigaciones Energeticas Medioambientales y Tecnologicas, E-28040 Madrid, Spain} 
\affiliation{Massachusetts Institute of Technology, Cambridge, Massachusetts 02139, USA} 
\affiliation{Institute of Particle Physics: McGill University, Montr\'{e}al, Qu\'{e}bec, Canada H3A~2T8; Simon Fraser University, Burnaby, British Columbia, Canada V5A~1S6; University of Toronto, Toronto, Ontario, Canada M5S~1A7; and TRIUMF, Vancouver, British Columbia, Canada V6T~2A3} 
\affiliation{University of Michigan, Ann Arbor, Michigan 48109, USA} 
\affiliation{Michigan State University, East Lansing, Michigan 48824, USA}
\affiliation{Institution for Theoretical and Experimental Physics, ITEP, Moscow 117259, Russia}
\affiliation{University of New Mexico, Albuquerque, New Mexico 87131, USA} 
\affiliation{Northwestern University, Evanston, Illinois 60208, USA} 
\affiliation{The Ohio State University, Columbus, Ohio 43210, USA} 
\affiliation{Okayama University, Okayama 700-8530, Japan} 
\affiliation{Osaka City University, Osaka 588, Japan} 
\affiliation{University of Oxford, Oxford OX1 3RH, United Kingdom} 
\affiliation{Istituto Nazionale di Fisica Nucleare, Sezione di Padova-Trento, $^{bb}$University of Padova, I-35131 Padova, Italy} 
\affiliation{LPNHE, Universite Pierre et Marie Curie/IN2P3-CNRS, UMR7585, Paris, F-75252 France} 
\affiliation{University of Pennsylvania, Philadelphia, Pennsylvania 19104, USA}
\affiliation{Istituto Nazionale di Fisica Nucleare Pisa, $^{cc}$University of Pisa, $^{dd}$University of Siena and $^{ee}$Scuola Normale Superiore, I-56127 Pisa, Italy} 
\affiliation{University of Pittsburgh, Pittsburgh, Pennsylvania 15260, USA} 
\affiliation{Purdue University, West Lafayette, Indiana 47907, USA} 
\affiliation{University of Rochester, Rochester, New York 14627, USA} 
\affiliation{The Rockefeller University, New York, New York 10065, USA} 
\affiliation{Istituto Nazionale di Fisica Nucleare, Sezione di Roma 1, $^{ff}$Sapienza Universit\`{a} di Roma, I-00185 Roma, Italy} 

\affiliation{Rutgers University, Piscataway, New Jersey 08855, USA} 
\affiliation{Texas A\&M University, College Station, Texas 77843, USA} 
\affiliation{Istituto Nazionale di Fisica Nucleare Trieste/Udine, I-34100 Trieste, $^{gg}$University of Udine, Udine, Italy} 
\affiliation{University of Tsukuba, Tsukuba, Ibaraki 305, Japan} 
\affiliation{Tufts University, Medford, Massachusetts 02155, USA} 
\affiliation{University of Virginia, Charlottesville, Virginia 22906, USA}
\affiliation{Waseda University, Tokyo 169, Japan} 
\affiliation{Wayne State University, Detroit, Michigan 48201, USA} 
\affiliation{University of Wisconsin, Madison, Wisconsin 53706, USA} 
\affiliation{Yale University, New Haven, Connecticut 06520, USA} 
\author{T.~Aaltonen}
\affiliation{Division of High Energy Physics, Department of Physics, University of Helsinki and Helsinki Institute of Physics, FIN-00014, Helsinki, Finland}
\author{B.~\'{A}lvarez~Gonz\'{a}lez$^w$}
\affiliation{Instituto de Fisica de Cantabria, CSIC-University of Cantabria, 39005 Santander, Spain}
\author{S.~Amerio}
\affiliation{Istituto Nazionale di Fisica Nucleare, Sezione di Padova-Trento, $^{bb}$University of Padova, I-35131 Padova, Italy} 

\author{D.~Amidei}
\affiliation{University of Michigan, Ann Arbor, Michigan 48109, USA}
\author{A.~Anastassov}
\affiliation{Northwestern University, Evanston, Illinois 60208, USA}
\author{A.~Annovi}
\affiliation{Laboratori Nazionali di Frascati, Istituto Nazionale di Fisica Nucleare, I-00044 Frascati, Italy}
\author{J.~Antos}
\affiliation{Comenius University, 842 48 Bratislava, Slovakia; Institute of Experimental Physics, 040 01 Kosice, Slovakia}
\author{G.~Apollinari}
\affiliation{Fermi National Accelerator Laboratory, Batavia, Illinois 60510, USA}
\author{J.A.~Appel}
\affiliation{Fermi National Accelerator Laboratory, Batavia, Illinois 60510, USA}
\author{A.~Apresyan}
\affiliation{Purdue University, West Lafayette, Indiana 47907, USA}
% ...
\collaboration{CDF Collaboration\footnote{With visitors from $^a$Istituto Nazionale di Fisica Nucleare, Sezione di Cagliari, 09042 Monserrato (Cagliari), Italy,
$^b$University of CA Irvine, Irvine, CA  92697, USA,
$^c$University of CA Santa Barbara, Santa Barbara, CA 93106, USA,
$^d$University of CA Santa Cruz, Santa Cruz, CA  95064, USA,
$^e$CERN,CH-1211 Geneva, Switzerland,
$^f$Cornell University, Ithaca, NY  14853, USA, 
$^g$University of Cyprus, Nicosia CY-1678, Cyprus, 
$^h$Office of Science, U.S. Department of Energy, Washington, DC 20585, USA,
$^i$University College Dublin, Dublin 4, Ireland,
$^j$University of Fukui, Fukui City, Fukui Prefecture, Japan 910-0017,
$^k$Universidad Iberoamericana, Mexico D.F., Mexico,
$^l$Iowa State University, Ames, IA  50011, USA,
$^m$University of Iowa, Iowa City, IA  52242, USA,
$^n$Kinki University, Higashi-Osaka City, Japan 577-8502,
$^o$Kansas State University, Manhattan, KS 66506, USA,
$^p$University of Manchester, Manchester M13 9PL, United Kingdom,
$^q$Queen Mary, University of London, London, E1 4NS, United Kingdom,
$^r$University of Melbourne, Victoria 3010, Australia,
$^s$Muons, Inc., Batavia, IL 60510, USA,
$^t$Nagasaki Institute of Applied Science, Nagasaki, Japan, 
$^u$National Research Nuclear University, Moscow, Russia,
$^v$University of Notre Dame, Notre Dame, IN 46556, USA,
$^w$Universidad de Oviedo, E-33007 Oviedo, Spain, 
$^x$Texas Tech University, Lubbock, TX  79609, USA,
$^y$Universidad Tecnica Federico Santa Maria, 110v Valparaiso, Chile,
$^z$Yarmouk University, Irbid 211-63, Jordan,
$^{hh}$On leave from J.~Stefan Institute, Ljubljana, Slovenia, 
}}
\noaffiliation

\collaboration{CDF Collaboration}
\noaffiliation

\date{\today}

\begin{abstract}
  We present a search for the standard model Higgs boson production in
  association with a $W$ boson in proton-antiproton collisions
  ($p\bar{p}\rightarrow W^\pm H \rightarrow \ell\nu b\bar{b}$) at a
  center of mass energy of 1.96 TeV. The search employs data collected
  with the CDF II detector which correspond to an integrated
  luminosity of approximately 2.7 fb$^{-1}$. We recorded this data 
  with two kinds of triggers. The first kind required high-p$_T$ charged
  leptons and the second required both missing transverse energy and jets.
  The search selects events consistent with a signature 
  of a single lepton ($e^\pm/\mu^\pm$),
  missing transverse energy, and two jets. Jets corresponding to
  bottom quarks are identified with a secondary vertex tagging method
  and a jet probability tagging method. Kinematic information is fed
  in an artificial neural network to improve discrimination between
  signal and background. The search finds that both the observed
  number of events and the neural network output distributions are
  consistent with the standard model background expectations, and sets
  95\% confidence level upper limits on the production cross
  section times branching ratio. The limits are expressed as a ratio
  to the standard model production rate. The limits range from 3.6
  (4.3 expected) to 61.1 (43.2 expected) for Higgs masses from 100 to
  150 GeV/$c^{2}$, respectively.

\end{abstract}


\maketitle

\section{Introduction}

Standard electroweak theory predicts the existence of a single
fundamental scalar particle, the Higgs boson, which arises as a result
of spontaneous electroweak symmetry breaking~\cite{Higgs:1964pj}. The
Higgs boson is the only fundamental standard model particle which has
not been experimentally observed.  Direct searches at LEP2 and the
Tevatron have yielded constraints on the Higgs boson mass.  LEP2 data
exclude a Higgs boson with $m_H < 114.4\,\mathrm{GeV}/c^2$ at 95\%
confidence level (C.L.).  Recently, the Tevatron has excluded at 95\%
C.L. the mass range $154 < m_H <
175\,\mathrm{GeV}/c^2$~\cite{TevHiggsComb2010}.  In addition, recent
global fits to electroweak data yielded a one-sided 95\% confidence
level upper limit of 158~$\mathrm{GeV}/c^2$~\cite{LEP_EWWG_2008}. If
the experimental lower limit of $114.4\,\mathrm{GeV}/c^2$ is included
in the fit, then the upper limit raises to 185~$\mathrm{GeV}/c^2$.




The Higgs boson branching ratios depend on the particle's mass.  If
the Higgs boson has a low mass ($m_H<135\,\mathrm{GeV}/c^2$), it decays
mostly to $b\bar{b}$~\cite{Djouadi:1997yw}. If the Higgs boson has a high
mass ($m_H>135\,\mathrm{GeV}/c^2$), then it preferentially decays to
$W^+W^-$. 

Higgs boson production in association with a $W$ boson ($WH$) is the
most sensitive low-mass search channel at the Tevatron. $WH$
production is more sensitive than $ZH$ production because it has a
larger cross section. It is more sensitive than direct Higgs
production $gg \rightarrow H \rightarrow b\bar{b}$ because it has a
smaller QCD background.

Searches for $WH\rightarrow \ell\nu b\bar{b}$ at
$\sqrt{s}=1.96\,\rm{TeV}$ have been recently reported by CDF using
1.9~fb$^{-1}$~\cite{WH2FB}, and D0 using 440
pb$^{-1}$~\cite{new_d0}. The CDF analysis looked for $WH$ production
in charged-lepton-triggered events. It improved on prior results by
employing a combination of different jet flavor identification
algorithms~\cite{Acosta:2004hw}. Flavor identification algorithms
distinguish between jets that are induced by light partons ($u,d,s,g$)
and jets containing the debris of heavy quarks ($b,c$). The analysis
also introduced multivariate techniques that use several kinematic
variables to distinguish signal from background. The analysis set
upper limits on the Higgs boson production rate, defined as the cross
section times branching ratio $\sigma \cdot {\cal B}$ for mass
hypotheses ranging from 110 to $150\,\mathrm{GeV}/c^2$.  The rate was
constrained to be less than 1.0 pb at 95\% C.L. for $m_H=110$ and less
than 1.2 pb for $150\,\mathrm{GeV}/c^2$. This corresponds to a limit
of 7.5 to 102 times the standard model cross section. More recently,
CDF has produced a search with 2.7~fb$^{-1}$ of data that combines
both neural network and matrix element techniques
~\cite{WH_COMBO_PRL}. The search we present here is an ingredient in
the most recent combination.


The new search for $WH\rightarrow \ell\nu b\bar{b}$ reported here
builds on the previous CDF result by adding more data and introducing
new analysis techniques for identifying $W$ candidate events that have
been recorded using triggers involving missing transverse energy \MET
and jets. We use 2.7 fb$^{-1}$ of data in our search, which is an
increase of nearly 50\% over the prior search.  Our analysis uses both
events recorded with a charged-lepton trigger and events recorded by a
trigger that selects missing transverse energy \MET and two jets. The
missing transverse energy vector is the negative of the vector sum of
calorimeter tower energy deposits in the event. It is corrected for
the transverse momentum of any muons in the event.  \MET is the
magnitude of the missing transverse energy vector. Missing transverse
energy suggests that a neutrino from a $W$ decay was present in an
event. We identify $W$ candidates in \MET + jet events using looser
charged-lepton identification requirements that recover muons that
fell into gaps in the muon system. We show that including these events
significantly increases the search sample and that these new events
have a purity that is comparable to the samples using charged-lepton
triggers samples.

We describe the analysis as follows: in Section~\ref{sec:detector} we
describe the CDF II detector. We explain the event selection criteria
in Sec.~\ref{sec:eventSelection}, focusing especially on the
identification of loose muons.  In Sec.~\ref{sec:btag} we
discuss the $b$-tagging algorithms.  We estimate contributions from
the standard model (SM) backgrounds and show the results in
Sec.~\ref{sec:bkg}. In Sec.~\ref{sec:Acceptance}, we estimate our
signal acceptance and systematic
uncertainties. Sec.~\ref{sec:neuralnet} describes the multivariate
technique that we use to enhance our discrimination of signal from
backgrounds. We report our measured limits in Sec.~\ref{sec:limit} and
interpret the result in Sec.~\ref{sec:conclusions}.

\section{CDF II Detector}
\label{sec:detector}
The CDF II detector~\cite{Acosta:2004yw} geometry is described using a
cylindrical coordinate system.  The $z$-axis follows the proton
direction, the azimuthal angle is $\phi$, and the polar angle $\theta$
is usually expressed through the pseudorapidity $\eta =
-\ln(\tan(\theta/2))$. The detector is approximately symmetric in
$\eta$ and about the $z$ axis.  The transverse energy is defined as
$E_T=E\sin \theta$ and transverse momentum as $p_T=p \sin \theta$.

Charged particles are tracked by a system of silicon microstrip
detectors and a large open cell drift chamber in the region
$|\eta|\leq 2.0$ and $|\eta|\leq 1.0$, respectively. The open cell
drift chamber is called the central outer tracker (COT).  The tracking
detectors are immersed in a $1.4\,\mathrm{T}$ solenoidal magnetic
field aligned coaxially with the incoming beams, allowing measurement
of charged particle momentum.

The transverse momentum resolution is measured to be $\delta p_T/p_T
\approx 0.1\% \cdot p_T$(GeV) for the combined tracking system.  The
track impact parameter $d_0$ is the distance from the event vertex to
the track's closest approach in the transverse plane. It has a
resolution of $\sigma(d_0) \approx 40\,\mu{\rm m}$ of which 30 ~$\mu{\rm m}$
is due to the size of the beam spot.

Outside of the tracking systems and the solenoid, segmented
calorimeters with projective tower geometry are used to reconstruct
electromagnetic and hadronic
showers~\cite{Balka:1987ty,Bertolucci:1987zn,Albrow:2001jw} over the
pseudorapidity range $|\eta|<3.6$.  A transverse energy is measured
in each calorimeter tower where $\theta$ is calculated using the
measured $z$ position of the event vertex and the tower location.

Small contiguous groups of calorimeter towers with energy deposits are
identified and summed together into an energy cluster.  Jets are
identified by summing energies deposited in electromagnetic (EM) and hadronic
calorimeter (HAD) towers that fall within a cone of radius $\Delta R =
\sqrt{(\Delta \phi^2 + \Delta \eta^2)} \leq 0.4$ units around a
high-$E_T$ seed cluster~\cite{Abe:1991ui}.  Jet energies are corrected
for calorimeter non-linearity, losses in the gaps between towers and
multiple primary interactions~\cite{Bhatti:2005ai}. Electron
candidates are identified in the central electromagnetic calorimeter
(CEM) as isolated, electromagnetic clusters that match a track in the
pseudorapidity range $|\eta|<1.1$.  The electron transverse energy is
reconstructed from the electromagnetic cluster with a precision
$\sigma(E_T)/E_T = 13.5\%/\sqrt{E_T/(\mathrm{GeV})} \oplus
2\%$~\cite{Balka:1987ty}.


This analysis uses three separate muon detectors and the gaps in
between the detectors to identify muon candidates.  After at least
five hadronic interaction lengths in the calorimeter, the muons
encounter the first set of four layers of planar drift chambers
(CMU). After passing through another 60~cm of steel, the muons reach
an additional four layers of planar drift chambers (CMP). Muons
require $p_T > 1.4\,{\rm GeV}/c$ to reach the CMU~\cite{Ascoli:1987av}
and an $p_T > 2.0$ GeV/$c$ to reach the CMP~\cite{Dorigo:2000ip}.  Muon candidates
are then identified as tracks that extrapolate to line segments or
``stubs'' in one of the muon detectors.  A track that is linked to
both CMU and CMP stubs is called a CMUP muon. These two systems cover
the same central pseudorapidity region with $|\eta| \leq 0.6$.  Muons
that exit the calorimeters at $ 0.6 \leq |\eta| \leq 1.0$ are detected
by the CMX system of four drift layers and are called CMX muons.
Tracks that point to a gap in the CMX or CMUP muon system are called
isolated track muon candidates.

The CDF trigger system is a three-level filter, with tracking
information available even at the first level~\cite{Thomson:2002xp}.
Events used in this analysis have passed either the electron trigger,
the muon trigger, or the missing transverse energy \MET trigger
selection.  The lepton trigger selection is identical to the selection
used in~\cite{WH2FB}.  The first stage of the central electron trigger
requires a track with $p_T > 8$~GeV/$c$ pointing to a tower with $E_T
> 8$~GeV and $E_{\mathrm{HAD}}/E_{\mathrm{EM}}<0.125$, where $E_{\mathrm{HAD}}$
is the hadronic calorimeter energy and $E_{\mathrm{EM}}$ is the 
electromagnetic calorimeter energy.
The first stage of the muon trigger requires a track with $p_T >
4$~GeV/$c$ (CMUP) or 8~GeV/$c$ (CMX) pointing to a muon stub.  For
lepton triggers, a complete lepton reconstruction is performed online
in the final trigger stage, where we require~$E_T$~$>$~18~GeV/c$^2$ 
for central electrons (CEM), and $p_T > 18\,\mathrm{GeV}/c$ for muons (CMUP,CMX).


The \MET plus two jets trigger has been previously used in the
$V(=W,Z)H \rightarrow$ \MET $+b\bar{b}$ Higgs search \cite{METBB_1fb}
and offers a chance to reconstruct $WH$ events that did not fire the
high-$p_T$ lepton trigger. The trigger's  requirements are two
jets and missing transverse energy. The two jets must have $E_T >
10$~GeV, and one must be in the central region $|\eta| < 0.9$.  The missing
transverse energy calculation  that is used in the trigger, $\METraw$, 
assumes that primary vertex of the event is at the center of the
detector and does not correct for muons.  The trigger requires
$\METraw > 35$~GeV.  Sections~\ref{sec:eventSelection}
and~\ref{sec:Acceptance} discuss the implications of these trigger
requirements on the event selection and trigger efficiency.

\section{Event Selection}
\label{sec:eventSelection}
The observable final state from $WH$ production and decay consists
of a high-$p_T$ lepton, missing transverse energy, and two jets.  This
section provides an overview of how we reconstruct and identify each
part of the $WH$ decay, focusing especially on isolated track
reconstruction, which is new for this result.  Additional details on
the event reconstruction can be found in Ref.~\cite{WH2FB}.

\subsection{Lepton Identification}

We use several different lepton identification algorithms in order to
include events from multiple trigger paths. Each algorithm requires a
single high-$p_T$ ($> 20 $~GeV/$c$), isolated charged lepton consistent with
leptonic $W$ boson decay.  We employ the same electron and muon
identification algorithms as the CDF $W$ cross section
measurement~\cite{Acosta:2004uq} and the prior CDF $WH$
search~\cite{WH2FB}. We classify the leptons according to the
sub-detector that recorded them: CEM electrons, CMUP muons, and CMX
muons. We supplement the lepton identification with an
additional category called ``isolated tracks''.  An isolated track event
is required to have a single, energetic track that is isolated from
other track activity in the event and that has not been reconstructed
as an electron or a muon using the other algorithms mentioned above.

\begin{figure} [htbp]
  \begin{center}
    \includegraphics[width=0.48\textwidth]{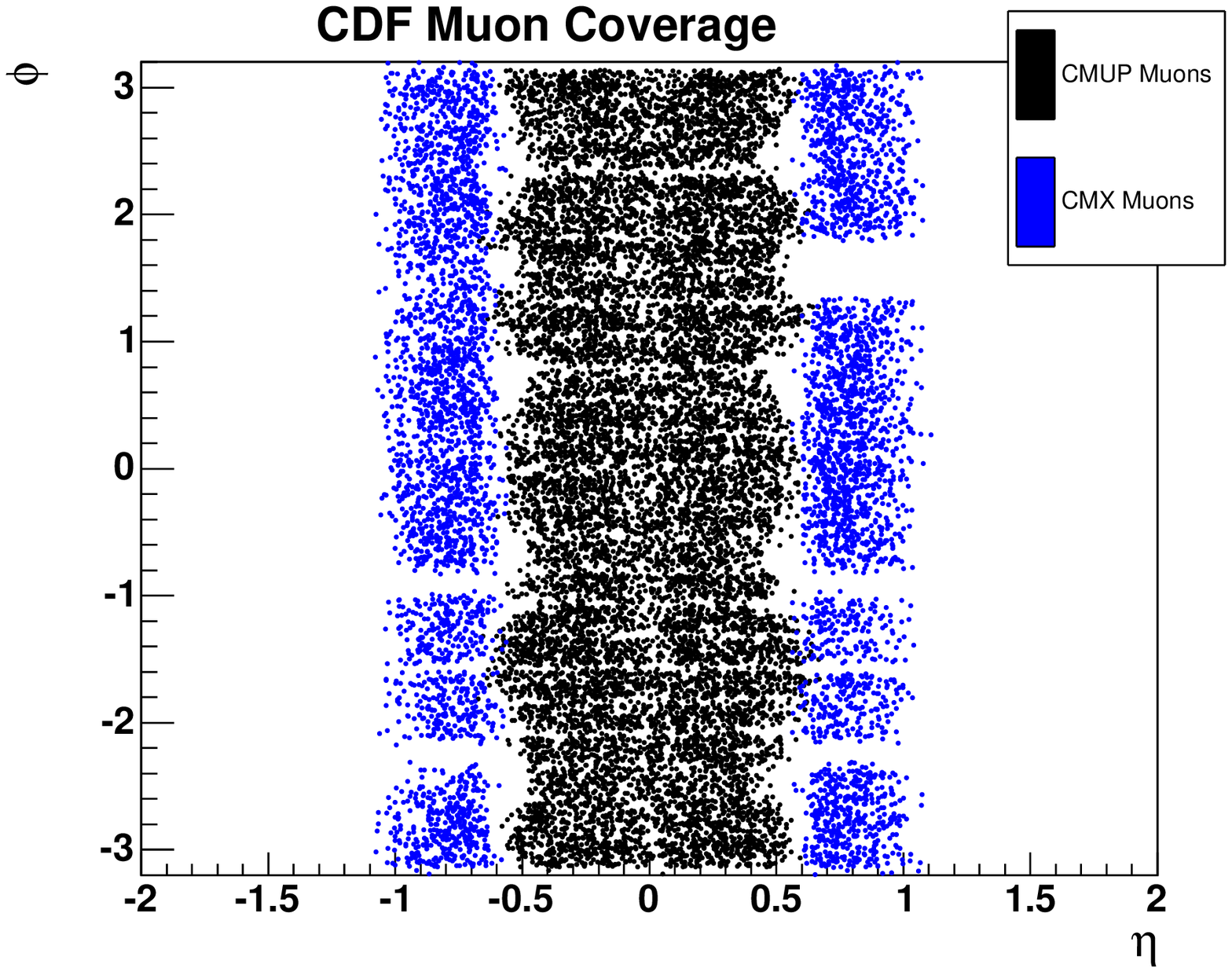}
    \includegraphics[width=0.48\textwidth]{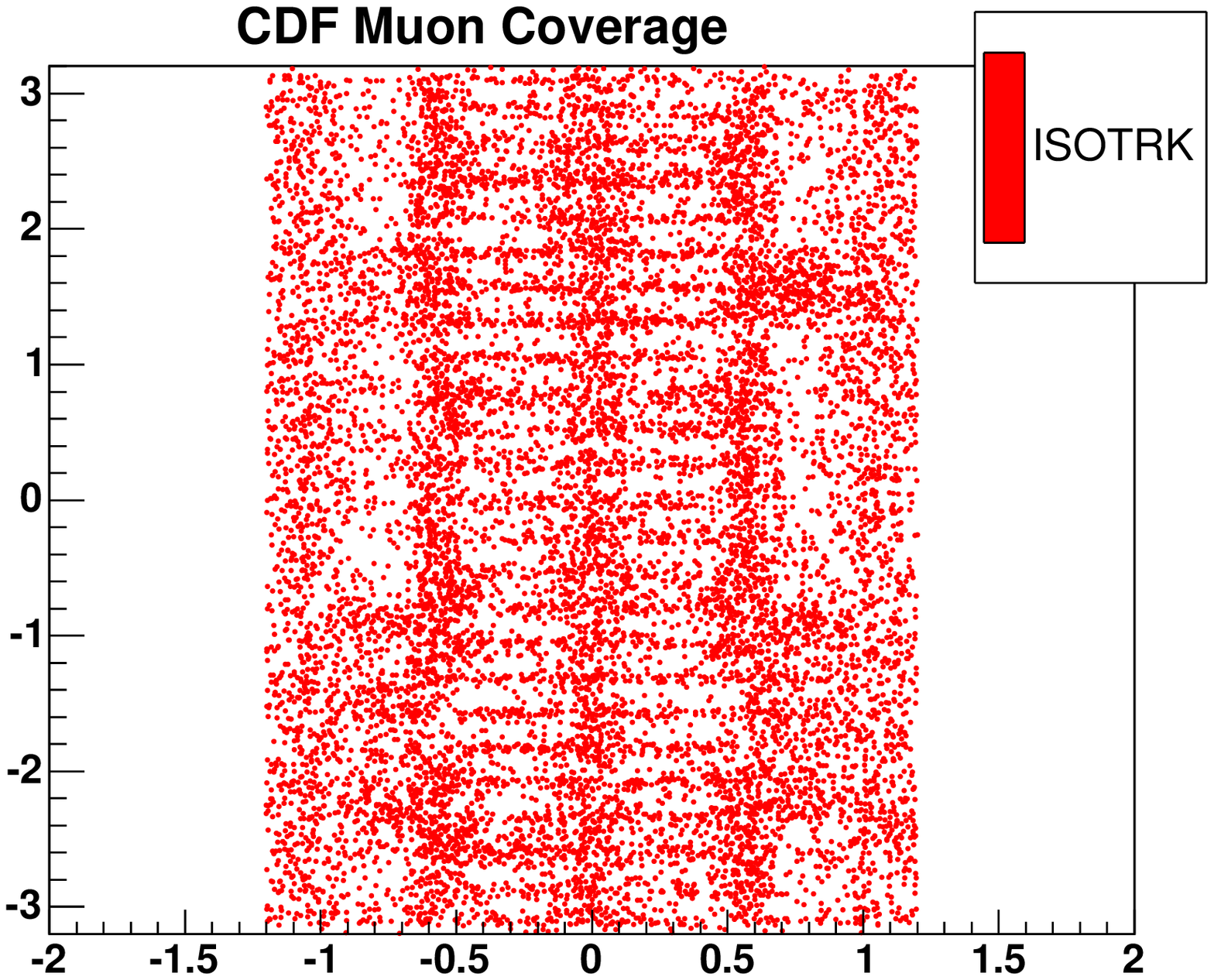}
  \end{center}

  \caption{(Left) Angular distribution of $WH$ Monte Carlo 
    muon triggered events.
    Note the cracks between CMUP chambers and the gap between
    the CMUP and CMX.
    (Right) Isolated track events  recover high-$p_T$ muons 
    that fall in the muon chamber gaps.
  }
  \label{fig:EtaPhi}
  
\end{figure}

The isolated track selection is designed to complement the trigger muon 
selection in that it finds muons that did not leave hits in the muon chambers, 
and therefore, could not have fired the muon trigger.  Figure \ref{fig:EtaPhi} shows how
isolated track events increase overall muon coverage. The isolated track events
are concentrated in the regions where there
is no other muon coverage. Including isolated track events increases
the acceptance by 25\% relative to the acceptance of charged-lepton triggers.

We identify isolated tracks based on criteria used in the top lepton
plus track cross section measurement \cite{TopXsecIsotrk}. Table
\ref{table:isotrk_cuts} outlines the specific isolated track selection
criteria. The track isolation variable quantifies the amount of track
activity near the lepton candidate. It is defined as
\begin{equation}
{\rm TrkIsol} = \frac{p_T(candidate)}{p_T(candidate) + \sum{p_T(trk)}}  ~~,
\label{eqn:trkisol}
\end{equation}
where $\sum{p_T(trk)}$ is the sum of the $p_T$ of tracks
that meet the requirements in Table \ref{tab:trkIsolReq}.
Using this definition, a track with no surrounding
activity has an isolation of 1.0. We require track
isolation to be $>$ 0.9.

We veto events with an identified charged lepton that fires the
trigger (CEM, CMUP, CMX) in order to ensure that the data sets are
disjoint.  In addition, we veto events with two or more isolated
tracks or a single isolated track that falls inside the cone of a jet
($\Delta R <$ 0.4), as these events are unlikely to have come from $W
\rightarrow \mu \nu$ decay.

\begin{table}
  \begin{center}
    \begin{tabular}{lc}
      \hline
      \hline
      Variable & Cut \\
      \hline
      $p_T$ & $>$ 20 GeV/$c$ \\
      $|z_0|$ &  $<$ 60 cm \\      
      $|d_0|$ & $<$ 0.02 cm\\
      $|d_0|$ (no Si hits) &  $<$ 0.2 cm\\
      track isolation  & $>$ 0.9 \\
      Axial COT hits & $\geq$ 24 \\
      Stereo COT Hits & $\geq$ 20 \\
      Num Si Hits (only if num expected hits $\geq 3$) & $\geq$ 3 \\
      \hline
      \hline
    \end{tabular}
  \end{center}
  \caption{Isolated track identification requirements. In the table, 
    $d_0$ is the track impact parameter, $d_0$ (no Si Hits) is the impact parameter 
    for tracks that have no silicon tracker hits, $z_0$ is position along the direction of the beamline 
    of the closest approach of the track to the beamline, and the Axial and Stereo hits are on tracks the open cell drift 
    chamber (COT). We 
    define track isolation according to equation \ref{eqn:trkisol}. }
  \label{table:isotrk_cuts}
\end{table}

\begin{table}
  \begin{center}
    \begin{tabular}{lc}
      \hline
      \hline
      Variable  &   Cut \\
      \hline
      $p_T$   & $>$ 0.5 GeV/$c$ \\
      $\Delta R$(trk, candidate) &  $<$ 0.4 \\
      $\Delta Z$(trk, candidate) & $<$ 5 cm \\
      Number of COT axial hits  &  $>$ 20 \\
      Number of COT stereo hits &  $>$ 10 \\
      \hline
      \hline
    \end{tabular}
  \end{center}
  \caption{   \label{tab:trkIsolReq} Requirements for tracks included in track isolation calculation.}

\end{table}

\subsection{Jet Selection}

$WH$ signal events have two high-$E_T$ jets from the $H \rightarrow
b\bar{b}$ decays. We define reconstructed jets using a cone of $\Delta
R < 0.4$, where $\Delta R = \sqrt{\Delta \phi^2 + \Delta \eta^2}$. We
require jets to have $E_T>20$~GeV and $|\eta|<2.0$.  The $\eta$ cut
ensures that the jets are within the fiducial volume of the silicon
detector.  The jet energies are corrected to account for variations in
calorimeter response in $\eta$, calorimeter non-linearity, and energy
from additional interactions in the same bunch crossing. Monte Carlo
simulations (MC) show that about 60\% of $WH$ events passing our
selections result in two-jet events.  The remainder is split evenly
between events with one or three jets.  Events with one or three jets
have a worse signal-to-background ratio than those with two jets due
to contamination from background processes such as $W$+jets and
$t\bar{t}$, respectively. We limit our search for $WH\rightarrow
\ell\nu b\bar b$ to events with $W$ + exactly two jets.

For events collected on the \MET + jets trigger, we require the jets
to have an $E_T > 25$~GeV to ensure that they are above the trigger
threshold.  One of the two jets must be in the central region $|\eta| <
0.9$ to match the requirements of the trigger. In addition, because
the trigger has a low efficiency for jets that are close together, we
require the jets to be well-separated ($\Delta R > 1.0$).

Table~\ref{tab:jetSelection} summarizes the jet selection criteria for 
events in each trigger sample.

\begin{table}
  \begin{center}
    \begin{tabular}{cc}
      \hline
      \hline
       Trigger Sample & Jet Selection \\
      \hline
       \multirow{2}{*}{Charged Leptons} & $E_T>20$~GeV \\
       & $|\eta|<2.0$ \\
       \hline
       \multirow{4}{*}{\MET + Jets} & $E_T > 25$~GeV \\ 
       & $|\eta|<2.0$  \\
       & At least one jet $|\eta| < 0.9$  \\
       & $\Delta R > 1.0$  \\
      \hline
      \hline
    \end{tabular}
  \end{center}

  \caption{Jet selection criteria for events in our different 
  trigger samples. \label{tab:jetSelection}}
\end{table}

In calculating event kinematics we find it useful to consider loose
jets that have either somewhat smaller $E_T$ than our 
cuts or have high-$E_T$ but are further forward than our standard
jets. We call these jets ``loose jets''. We do not use them directly 
in our event selection, but we do use them in calculating kinematic
variables. We define loose jets to be jets with $E_T > 12$~GeV in the
region $|\eta| < 2.0$, and $E_T > 20$~GeV in the region $2.0 < |\eta|
< 2.4$.

\subsection{Missing Transverse Energy}

The presence of a neutrino from the $W$ decay is inferred from the
presence of a significant amount of missing transverse energy.  The
missing transverse energy vector is the negative of the vector sum of
all calorimeter tower energy deposits with $|\eta| < 3.6$. The \MET is
the magnitude of the missing $E_T$ vector.  We correct the energy of
jets in the event~\cite{Bhatti:2005ai} and propagate the corrections
to the \MET. We also account for the momentum of any high $p_T$ muons.
When we calculate \MET, we use $z$-position of the primary vertex to
get the correct $E_{T}$ for each calorimeter tower. Some events have
more than one vertex. In this case, We use the sum of the transverse
momentum of the tracks associated with each vertex to distinguish
between the vertexes. The primary vertex is the one with the highest
sum of the track transverse momentum.  We then require \MET to exceed
$20$~GeV.

\subsection{$b$-jet identification}
\label{sec:btag}

Both of the jets in $WH$ events originate from $H\rightarrow b\bar{b}$
decays. Many backgrounds have jets that come from light-flavor partons
($u,d,c,s,g$), such as $W$ + jets and QCD. Jets from $b$ quarks can be
distinguished from light-flavor jets by looking for the decay of
long-lived $B$ hadrons. We use the same $b$-jet identification strategy
as the previous $WH$ search \cite{WH2FB}. We employ two separate
algorithms to identify $B$ hadrons. The secondary vertex tagging
algorithm~\cite{Abulencia:2006in} takes tracks within a
jet and attempts to reconstruct a secondary vertex. If a vertex is
found and it is significantly displaced from the primary vertex, 
the jet is identified, or tagged, as a $b$ jet. The Jet Probability
algorithm \cite{Abulencia:2006kv} also uses tracking information
inside of jets to identify $B$ decays. Instead of requiring a
secondary vertex, the algorithm looks at the distribution of impact
parameters for tracks inside a jet. If the jet has a significant
number of large impact parameter tracks, then it is tagged as a
$b$-jet. Jet probability tags have a lower purity than secondary
vertex tags.



\subsection{Lepton + Jets Selection}

After identifying the final state objects in the event, we purify the
sample with quality cuts. We fit a subset of well-measured tracks
coming from the beamline to determine the event's primary vertex. The
longitudinal coordinate $z_0$ of the lepton track's point of closest
approach to the beamline must be within 5~cm of the primary vertex to
ensure that the lepton and the jets come from the same hard
interaction. We reduce backgrounds from $Z$ boson decays by vetoing
events where the invariant mass of the lepton and a second track with
$p_T>10\,\mathrm{GeV}/c$ falls in the $Z$-boson mass window $76<
m_{\ell-trk} < 106\,\mathrm{GeV}/c^2$.

We use the b-jet tagging strategy developed in the previous $WH$ search
\cite{WH2FB}. We require at least one jet to be $b$-tagged with the
secondary vertex algorithm, and then we divide our sample into three
exclusive categories of varying purity.  Events with two secondary
vertex tagged jets have the highest purity, followed by events with one secondary
vertex tagged jet and one jet probability tagged jet. In the lowest 
purity events, there is only one secondary vertex tagged jet.

We further purify the sample with exactly one secondary vertex tagged
jet by using kinematic and angular cuts designed to reject QCD events
with fake $W$ signatures.  The kinematics of the QCD contamination
vary with the lepton signature they mimic. We therefore apply a
separate veto to each lepton subsample. 

One approach we use to reduce QCD is to cut on a variable
correlated with mismeasurement. The observation of single top quark
production \cite{ST_OBSV_PRD} demonstrated that missing transverse
energy significance \METsig is a useful variable to remove QCD
contamination.  Missing transverse energy significance \METsig quantifies the
likelihood that the measured \MET comes from jet
mismeasurements. \METsig is defined as follows:
\begin{equation}
S_{/\!\!\!\!\!E_{T}} = \frac{\METtext}{(\sum_{jets}C_{JES}^{2}\cos^{2}(\Delta\phi_{\METtext, jet})E_{T,jet}^{raw}+\cos^{2}(\Delta\phi_{\METtext, uncl })E_{T,uncl})^{1/2}} ~,
\label{eqMETsig}
\end{equation}
where $C_{JES}$ is the jet energy correction factor,
$\Delta\phi_{\METtext, jet}$ is the azimuthal angle between the jet
and the \MET direction, $E_{T,jet}^{raw}$ is the uncorrected jet
$E_{T}$, unclustered energy is energy not associated with a jet,
$E_{T,uncl}$ is the transverse unclustered energy, and
$\Delta\phi_{\METtext, uncl}$ is the azimuthal angle between the
unclustered energy direction and the \MET direction. The lower the
value of \METsig, the more likely it is that the \MET comes from
fluctuations in jet energy measurements.
The uncertainty on the calorimeter energy not
clustered into one of the jets is also included.

Another useful approach for rejecting QCD backgrounds is to require
that the lepton momentum and \MET be consistent with the decay of a
$W$ boson.  However, since only the transverse component of the
neutrino momentum is available via \MET, the $W$ invariant mass cannot
be calculated.  Instead, if we ignore the neutrino $p_z$, we can
calculate the transverse mass as follows:

\begin{equation}
M_{T} = \sqrt{2 (p_{T}^{lep}E_{T}\!\!\!\!\!\!/-{\bm p_{T}}^{lep}\cdot{\bm E_{T}\!\!\!\!\!\!/})} 
\end {equation}

We use both $M_{T}$ and \METsig to remove QCD events from our sample.
Table~\ref{tab:qcdVetos} lists the different QCD veto cuts for each 
lepton type. The cuts were chosen to have high efficiency for 
events with a $W$ boson while rejecting the maximum amount of QCD
and minimizing disagreement between data and MC in the pretag sample.

\begin{table}
  \begin{center}
    \begin{tabular}{cc}
      \hline
      \hline
      Quantity & Cut \\
      \hline
      \multicolumn{2}{c}{CEM} \\
      \hline
		 $M_T$  & $ > 20$~GeV \\
		 \METsig & $ \geq -0.05 \cdot M_T + 3.5$ \\
		 \METsig & $ \geq 2.5 - 3.125 \cdot \Delta\phi_{MET,jet2}$ \\
      \hline
	  \multicolumn{2}{c}{CMUP,CMX} \\
      \hline
	     $M_T$ & $ > 10$~GeV \\
      \hline
      \multicolumn{2}{c}{ISOTRK} \\
      \hline
         $M_{T}$ & $> 10$ GeV \\
      \hline
      \hline
    \end{tabular}
  \end{center}

  \caption{QCD veto cuts for each lepton category. These cuts
  are applied to events with exactly one identified b-jet.
  \label{tab:qcdVetos}}
\end{table}



%

\section{Backgrounds}\label{sec:bkg}

The signature of $WH$ associated production is shared by a number of
processes that can produce the combination $\ell\nu b\bar b$. The
dominant backgrounds are $W$+jets production, $t\bar t$ production,
single top production, and QCD multijet production. Diboson production
and Z+jets production, collectively referred to as ``electroweak
backgrounds,'' contribute to the sample at smaller rates than any of
the other backgrounds.  Diboson production has a small contribution
because of its small cross section and, in the case of WW, lack of
$b$-jets at leading order. $Z$+jets production has a small
contribution because it has a small overlap with our single lepton
final state. Our estimate of the background rates uses a combination
of Monte Carlo techniques and data-driven estimates. Our data-driven
estimates use background-enriched control regions outside of our
search region to determine background properties.  We extrapolate the
background properties from the control regions to the search region
and assess an uncertainty on the estimates.  Our background techniques
are common to top cross section measurements~\cite{Abulencia:2006in},
single top searches~\cite{SingleTop_2fb}, and prior $WH$
searches~\cite{Aaltonen:2007wx}.  We provide an overview of the
background estimate below and discuss the details of each background
in the subsections that follow.

We first describe our background estimate for the sample of $\ell \nu
j j$ events without any tagging requirements applied, which we refer
to as the pretag sample.  This sample is composed of events from two classes
of processes: (1) events containing a high-$p_T$ lepton from a real
$W$ decay and (2) events in which the lepton is from a source other
than a $W$.  In the second class of events, referred to as QCD
multijet events, the high-$p_T$ lepton comes either from a jet that
fakes a lepton signature or from a real lepton produced in a
heavy-flavor decay. After the QCD multijet background is subtracted
off, what remains are events from a collection of processes that
include the production of a $W$ boson: primarily $W$ + jets
production, top production, and other electroweak backgrounds.  We use
a Monte Carlo based technique to estimate the relative contributions
of processes whose rates and topologies are described well by
next-to-leading order (NLO) calculations. These processes include
$t\bar{t}$, single top and diboson, and $Z$ + jets production. We
estimate their expected contribution $N$ using the theoretical NLO
cross section $\sigma$, Monte Carlo event detection efficiency
corrected to match the efficiency in the data $\epsilon$, and the
integrated luminosity of our dataset $\mathcal{L}_{int}$:
\begin{eqnarray}
  N = \sigma \cdot \epsilon \cdot \mathcal{L}_{int} 
\label{eqn:nEvents}
\end{eqnarray}
We subtract the contribution of these processes from the total number
of observed events.  After accounting both for the fraction of QCD multijet events
and for the top and other electroweak processes, what remains are the  pretag $W$+jets events,
whose contribution is estimated as follows:
\begin{eqnarray}
 N_{W+Jets}^{Pretag} = N_{Pretag}
 \cdot (1 - F_{QCD}) - N_{EWK} - N_{TOP} 
\end{eqnarray}
where $N_{Pretag}$ is the observed number of $\ell\nu j j$ pretag
events, $N_{EWK}$ is the number of estimated electroweak events, and
$N_{TOP}$ is the number of estimated top events. 

We estimate the number of tagged $W$ + jets events using the number of
pretag $W$ + jet events and a tag probability. We measure the tag
probabilities for both light and heavy-flavor jets in inclusive jet
data. The tag probability for heavy-flavor jets is $\epsilon_{tag}$,
and the tag probability for falsely tagged jets, called ``mistags'',
is $\epsilon_{mistag}$. $W+b\bar{b}$, $W+c\bar{c}$, and $W+cq$
production are collectively referred to as $W$ + heavy-flavor
processes. All other $W$ + jets production is referred to as $W$ +
light flavor. We use a $b$-tag scale factor to correct the Monte Carlo
tagging efficiency according to the tag efficiency observed in
data. We must estimate the fraction of $W$ + jet events that are $W$ +
heavy-flavor events $F_{HF}$ in our sample in order to use the
appropriate tag probabilities. We use $W$ + 1 jet data to calibrate
the heavy-flavor fraction from the Monte Carlo. We use the ratio of
the heavy-flavor fraction in the data $F^{data}_{HF}$ to the
heavy-flavor fraction in the Monte Carlo $F^{MC}_{HF}$ to calculate a
correction factor $K = F^{data}_{HF} / F^{MC}_{HF} $. We apply the
correction factor to the number of $W$ + heavy jets estimated with the
Monte Carlo.  After including this calibration, the number of $W$+jets
in the tagged sample is:

\begin{eqnarray}
  N_{W+HF}^{tagged} = N_{W+jets}^{pretag} \cdot
  (F_{HF} \cdot K) \cdot \epsilon_{tag}
\label{eqn:numWHeavy}
\end{eqnarray}

\begin{eqnarray}
  N_{W+LF}^{tagged} = N_{W+jets}^{pretag} \cdot
  (1 - F_{HF} \cdot K) \cdot \epsilon_{mistag}  
\label{eqn:numMistags}
\end{eqnarray}

The estimation of the rate of these backgrounds are done separately
for each jet bin in the data.  Below we describe the estimation of the
individual pieces in greater detail.



\subsection{Top and Electroweak Backgrounds}
\label{sec:TOPEWK}

The normalization of the diboson, $Z$+jets, top-pair, and single-top
backgrounds are based on the theoretical cross
sections~\cite{Campbell:2002tg,Acosta:2004uq,Cacciari:2003fi,Harris:2002md}
listed in Table~\ref{tbl:xsec}. The estimate from theory is
well-motivated because the cross sections for most of the processes
have small theoretical uncertainties. $Z$+jets is the only process
where the large corrections to the leading order process give large
uncertainties to the theoretical cross section. The impact of the
large uncertainty on our sensitivity is marginalized by the small
overlap of $Z$+jets with the $W$+jets final state. The background
contributions are estimated using the theory cross sections,
luminosity, and the Monte Carlo acceptance and $b$-tagging
efficiency. The Monte Carlo acceptance is corrected for lepton
identification, trigger efficiencies, and the $z$ vertex cut. We also
use a b-tagging scale factor to correct for the difference in tagging
efficiency in Monte Carlo compared to data.

%
%
%

\begin{table}
  \begin{center}
    \begin{tabular}{cc}
      \hline
      \hline 
      Process & Theoretical Cross Section \\ 
      \hline  
	$WW$ & 12.40 $\pm$ 0.80 pb \\ 
	$WZ$ & 3.96 $\pm$ 0.06 pb \\ 
	$ZZ$ & 1.58 $\pm$ 0.05 pb \\ 
	Single top $s$-channel & 0.88 $\pm$ 0.11 pb\\ 
	Single top $t$-channel & 1.98 $\pm$ 0.25 pb\\
    $t\bar{t}$ & 6.7 $\pm$ 0.83 pb \\
    $Z$ + Jets & 787.4 $\pm$ 85 pb \\
    \hline
    \hline
    \end{tabular}
    \caption{Theoretical
    cross sections~\cite{Campbell:2002tg,Acosta:2004uq,Cacciari:2003fi,Harris:2002md} and 
    uncertainties for the electroweak and  top
    backgrounds. Top cross sections assume a mass of 
    $m_t = 175\,\mathrm{GeV}/c^2$. 
    \label{tbl:xsec}}
  \end{center}
\end{table}

\subsection{QCD Multijet}
\label{sec:nonW}


QCD multijet events can fake a $W$ signature when a jet fakes a lepton
and overall mismeasurement leads to fake \MET.  Since these events do
not have real $W$ bosons in them, we also use the term non-W to
refer to QCD multijet events. It is difficult to identify the precise
sources of mismeasurement and handle them appropriately in a detector
simulation. The difficulty is increased by the large number of
processes that contribute to the composition of the QCD background at
unknown relative rates.  Each lepton category is susceptible to
different kinds of fakes. We use different QCD models for
central-lepton triggered events and isolated track events.  

We model central-lepton triggered QCD events using events where a jet
fired the electron trigger, passed the electron kinematic cuts, but
failed exactly two of the calorimeter or tracking quality cuts. Events
that fail these cuts will have the kinematic properties of $W$ events,
including isolation, but the sample will be enriched in fakes.  This
is the same model used in the CDF observation of single
top~\cite{ST_OBSV_PRD}.  As noted in that paper, these fake events
have the remarkable property that they model both electron and muon
fakes.

We model QCD events that fake an isolated track by using events
recorded on the \MET + 2 Jets trigger. We use events with muon
candidates that are not calorimeter isolated and are within the isolated
track acceptance ($|\eta| < 1.2$). Calorimeter isolation is defined as
the fraction of the lepton energy in a cone of $\Delta R = 0.4$ surrounding
the lepton. Non-isolated leptons are unlikely to come
from the decay of an on-shell $W$, and thus are enriched in fakes.

\begin{figure}[htbp]
  \begin{center}
    \includegraphics*[width=0.7\textwidth]{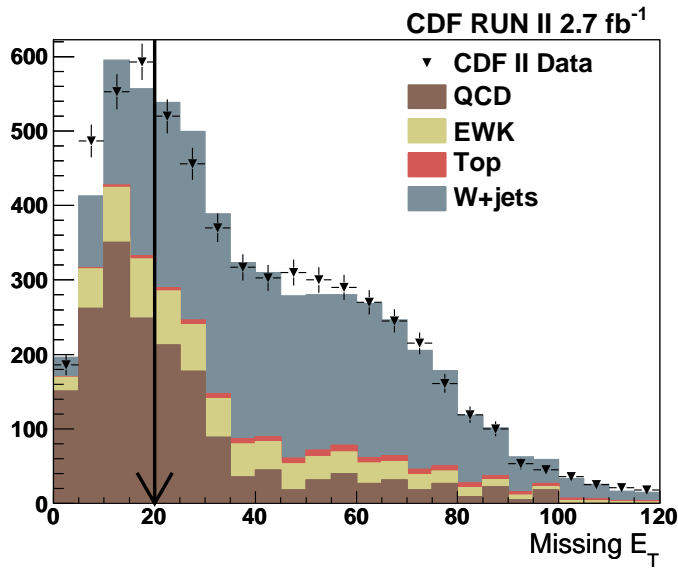}    
    \caption{Fit of the pretag isolated track \MET control region that
      is used to determine the QCD fraction of isolated track events. The
      arrow illustrates the \MET cut. We estimate a QCD fraction of
      19\% for the region with \MET $>$ 20 GeV. There is some
      disagreement between the data and our model in the low-\MET
      control region, and also around 50-55~GeV. The figure shows 
      just one QCD model. The difference between this nominal model
      and are alternate covers the modelling difference shown here. 
      We use the difference between the two models as our systematic 
      uncertainty. }
    \label{fig:isotrk_pretag_qcd_fit}
  \end{center}
\end{figure}

We estimate the amount of QCD background in each sample by fitting the
\MET spectrum in data. The fit includes the control region \MET $<$ 20
GeV, which is enriched in QCD fakes. Figure
\ref{fig:isotrk_pretag_qcd_fit} shows the \MET fit for isolated track
pretag events. The fit has one component with fixed normalization and
two templates whose normalizations can vary. The fixed component is a
combination of top and electroweak processes whose normalizations are
described in Section \ref{sec:TOPEWK}. We let the $W$ + jets template
vary along with the QCD template because there is a large uncertainty
on the $W$+jets cross section.  The QCD template has a \MET spectrum
that peaks near low \MET, and its normalization is driven by the low
\MET bins. The normalization of the $W$+jets template is driven by the
high \MET region. The fit determines the relative amounts of QCD and
$W$+jets in the full \MET sample, and we use these fit results to
determine the QCD fraction in the search region (\MET $>$ 20 GeV). For
isolated track events with two jets and no b-tag requirement, we
estimate a 19\% QCD fraction in the signal region, as shown in Fig.
\ref{fig:isotrk_pretag_qcd_fit}. The pretag QCD fractions for the
other lepton types are less than the isolated track fractions. Pretag
CEM electrons events have 10\% QCD fraction, and both CMUP and CMX
muon events have a 3\% QCD fraction. While isolated tracks have a
larger amount of QCD events than the other lepton types, the vast
majority of the isolated track events (81\%) still contain $W$ bosons.
We use the QCD fractions for each lepton type and tag category in the
calculations for the background summaries in Tables
\ref{tbl:m2_tlep_eq1tag} through \ref{tbl:m2_isotrk_gr2tag}.

We estimate the uncertainty of the QCD normalization by studying the
change in the QCD fraction due to changes in the QCD model. For tight
lepton events we use an alternate QCD model based on leptons that fail
our isolation requirements. We find a 40\% uncertainty
to the QCD normalization that covers the effect of using this
alternative model. We use the same uncertainty estimate for both
tight leptons and isolated tracks.

\subsection{$W$+Heavy-Flavor}
\label{sec:wPlusHeavy}
The number of $W$ + heavy flavor events is a fraction the number of
$W$ + light flavor events, as described by $F_{HF}$ in Equations
\ref{eqn:numWHeavy} and \ref{eqn:numMistags}.  The fraction of
$W$+heavy-flavor events has been studied extensively and is modeled in
the ALPGEN Monte Carlo
Generator~\cite{Mangano:2002ea,Corcella:2001wc}. We calibrate the
ALPGEN Version 2 $W$ + jets Monte Carlo heavy-flavor fraction to match
the observed heavy flavor fraction in the $W$ + 1 jet control region.
We use the same calibration of the heavy-flavor fraction as the single
top observation \cite{ST_OBSV_PRD}.  The calibration uses template
fits of flavor-separating variables in $b$-tagged $W$ + 1 jet data to
measure the heavy flavor fraction. The calibration measures $K$, the
calibration factor as defined in equation~\ref{eqn:numWHeavy}, to be
$K = 1.4 \pm 0.4$.

We can estimate the amount of $W$ + heavy flavor events in our signal
region by calculating the efficiency for these events to pass our tag 
requirements  $\epsilon_{tag}$. The efficiency $\epsilon_{tag}$ is
\begin{eqnarray}
   \epsilon_{tag} = 1 - \prod_{i}^{jets} (1 - p_{tag}^{i}),
   \label{eqn:tagEfficiency}
\end{eqnarray}
where $p_{tag}^{i}$ is the probability for jet $i$ in the event to
have a $b$-tag. The probability for a $b$-tagged Monte Carlo jet
originating from a $b$ or $c$ quark to have a $b$-tag in the data is
the $b$-tag scale factor. The $b$-tag scale factor is the ratio of
data to Monte Carlo $b$-tag efficiencies. It is estimated to be $0.95
\pm 0.04$ for secondary vertex tags \cite{Acosta:2004hw} and $0.85 \pm
0.07$ for jet probability tags \cite{Abulencia:2006kv}. In the case
where there are additional light-flavor jets produced in the $W$ +
heavy flavor events, there is a small chance for those light-flavor
jets to be incorrectly tagged as $b$-jets. We account for this in
Equation~\ref{eqn:tagEfficiency} by giving these just a small
probability to be incorrectly tagged. We call the probability 
to be incorrectly tagged the mistag probability. It is discussed in
detail in Section~\ref{sec:mistags}.





Table~\ref{tbl:HFfrac} shows the corrected heavy-flavor fractions for
our $W$ + heavy-flavor samples divided according to the 
heavy-flavor process and number of reconstructed jets. It 
is necessary to divide the samples by heavy-flavor
process because $b$- and $c$-jets have different tagging 
efficiencies. Table~\ref{tbl:HFtag} shows the corrected 
per-event tagging efficiencies. We calculate the W + heavy-flavor
normalizations using Eq.~\ref{eqn:numWHeavy} and the fractions
and efficiencies from the tables.

The two sources of uncertainties for the $W$ + heavy-flavor
backgrounds are the $b$-tag scale factor uncertainty and the heavy
flavor fraction uncertainty. We accommodate the $b$-tag scale factor
uncertainty by shifting the scale factor by $\pm 1\sigma$, propagating
the change through our background calculation, and using difference
between the shifted and nominal calculation as our error. We add this
error in quadrature with the heavy-flavor fraction uncertainty and 
use the total error as a constraint on the background in
our likelihood fit.

\begin{table}
  \begin{center}
    \begin{tabular}{cccccc}
      \hline
      \hline 
      \multicolumn{6}{c}{Corrected Heavy Flavor (HF) fraction (\%) } \\
      \multicolumn{6}{c}{of inclusive $W$ + jet events by jet multiplicity} \\
      \hline
      Process & Number of  Jets & \multicolumn{4}{c}{Fraction of Events by Jet Multiplicity} \\
              & matched to HF   & $W$ + 2 jets & $W$ + 3 jets & $W$ + 4 jets & $W$ + 5 jets \\
      \hline
      
      $Wb\bar b$ & (1$b$)  &  2.2   $\pm$ 0.88  &  3.5 $\pm$ 1.4 &   4.63 $\pm$ 1.8  &    5.5 $\pm$ 2.2 \\
      $Wb\bar b$ & (2$b$)  &  1.32  $\pm$ 0.52  &  2.6 $\pm$ 1.0  &   4.17 $\pm$ 1.7  &   6.0 $\pm$ 2.4 \\
      $Wc\bar c$ & (1$c$)  &  11 $\pm$ 4.4  &  14 $\pm$ 5.6  &   15.18 $\pm$  6.1    &   15.8 $\pm$ 6.3 \\
      $Wc\bar c$ & (2$c$)  &  2.1  $\pm$ 0.84  &  4.7 $\pm$ 1.9  &   7.69 $\pm$ 3.1  &  10.9 $\pm$  4.4 \\
      
      \hline
      \hline 
    \end{tabular}
    \caption{The corrected fraction of inclusive $W$ + jet events that
      contain heavy-flavor. The fractions are divided into separate
      categories according to the Monte Carlo flavor information for
      jets in the event and the number of reconstructed heavy-flavor
      jets.  For example, $Wb \bar b$ (1$b$) events have two
      $b$-quarks at the generator level, but only one $b$-quark
      matched to a reconstructed jet. The fractions from {\sc alpgen}
      Monte Carlo have been scaled by the data-derived calibration
      factor of $1.4\pm 0.4$.}
    \label{tbl:HFfrac}    
  \end{center}  
\end{table}

\begin{table}
  \begin{center}
    \begin{tabular}{ccccc}
      \hline
      \hline 
      \multicolumn{5}{c}{Corrected Per-event $b$-tag efficiencies} \\     
      \hline
      \multicolumn{5}{c}{One SECVTX Tag Efficiency} \\
      \hline
      Jet Multiplicity  &  2 jets & 3 jets & 4 jets & 5 jets \\
      \hline
      Event Eff (1$b$) (\%) &  23.10  &  24.68 &  25.02 &  27.14 \\
      Event Eff (2$b$) (\%) &  30.09  &  30.34 &  30.35 &  29.71 \\
      Event Eff (1$c$) (\%) &  7.02   &  7.69  &  8.68  &  10.24 \\
      Event Eff (2$c$) (\%) &  9.46   &  10.46 &  11.24 &  12.12 \\     
      \hline
      \multicolumn{5}{c}{Two SECVTX Tag Efficiency}   \\   
      \hline
      Jet Multiplicity  &  2 jets & 3 jets & 4 jets & 5 jets \\
      \hline
      Event Eff (1$b$) (\%)&   0.30  &   0.78 &   1.34  &  1.76  \\
      Event Eff (2$b$) (\%)&   8.76  &   9.68 &   10.18 &  11.14 \\
      Event Eff (1$c$) (\%)&   0.04  &   0.12 &   0.24  &  0.40  \\
      Event Eff (2$c$) (\%)&   0.38  &   0.55 &   0.88  &  0.91  \\

      \hline
      \multicolumn{5}{c}{One SECVTX TAG + One JETPROB Tag Efficiency} \\
      \hline
      Jet Multiplicity  &  2 jets & 3 jets & 4 jets & 5 jets \\
      \hline
      Event Eff (1$b$) (\%)&   0.79  &   1.75  &  2.57  &  3.74 \\
      Event Eff (2$b$) (\%)&   6.95  &   7.78  &  8.86  &  9.77 \\ 
      Event Eff (1$c$) (\%)&   0.20  &   0.47  &  0.78  &  1.24 \\
      Event Eff (2$c$) (\%)&   1.19  &   1.59  &  2.14  &  2.43 \\

      \hline
      \hline
      
    \end{tabular}
    \caption{The corrected per-event tagging efficiencies for events
      with heavy-flavor content. The event efficiencies are divided
      into separate categories depending on the Monte Carlo truth
      flavor information for jets in the event: 1$b$ events have one
      jet matched to $b$-quark, 2$b$ events have two jets matched to a
      $b$-quark, 1$c$ events have one jet matched to a $c$-quark, and
      2$c$ events have two jets matched to a $c$-quark.}
    \label{tbl:HFtag}    
  \end{center}  
\end{table}

\subsection{Mistagged Jets}
\label{sec:mistags}
$W$ + light flavor events with a fake $b$-tag migrate into our signal
region. Our estimate of the number of falsely tagged $W$+light flavor
events is based on the pretag number of $W$ + light flavor events and
the sample mistag probability $\epsilon_{mistag}$ in equation
\ref{eqn:numMistags}.  The sample mistag probability is based on the per-jet
mistag probability. For each event in our $W$ + light flavor Monte Carlo
samples, we apply the per-jet mistag probability to each jet and combine the
probabilities to get an event mistag probability.  We combine the event mistag
rates to get $\epsilon_{mistag}$. 

We estimate the per-jet mistag probability for each of our two tagging
algorithms using a data sample of generic jets with at least two
well-measured silicon tracks.  The decay length is defined as the
distance between the secondary vertex and the primary vertex in the
plane perpendicular to the beam direction.  This decay length is
signed based on whether the tracks are consistent with the decay of a
particle that was moving away from (positive sign) or towards
(negative sign) the primary vertex. False tags are
equally likely to have positive or negative decay lengths to first
order. The symmetry allows calibration of the false tag probability
using negative tags. There is a slightly greater chance for a false
tag to have a positive decay length due to material interaction, and
our estimate accounts for this asymmetry.  The false tag probability
for {\sc secvtx} is parameterized in bins of $\eta$, number of
vertices, jet $E_T$, track multiplicity, and the scalar sum of the
total event $E_T$ \cite{Abulencia:2006in}. We parameterize 
jet probability mistaging in jet $\eta$, $z$ position of primary
vertex, jet $E_{T}$, track multiplicity, and scalar sum of the total
event $E_{T}$.

We estimate the uncertainties on the per-jet mistag probability by using
negatively tagged jets in the data. The uncertainty estimates check
for consistency between the number of expected and observed negative
tags. The uncertainties are accounted for in the analysis by
fluctuating the per-jet tag probabilities by $\pm 1 \sigma$, and propagating the
change through the background estimate.

\subsection{Summary of Background Estimate}

\begin{figure}[htbp]
  \begin{center}
    \includegraphics*[width=0.9\textwidth]{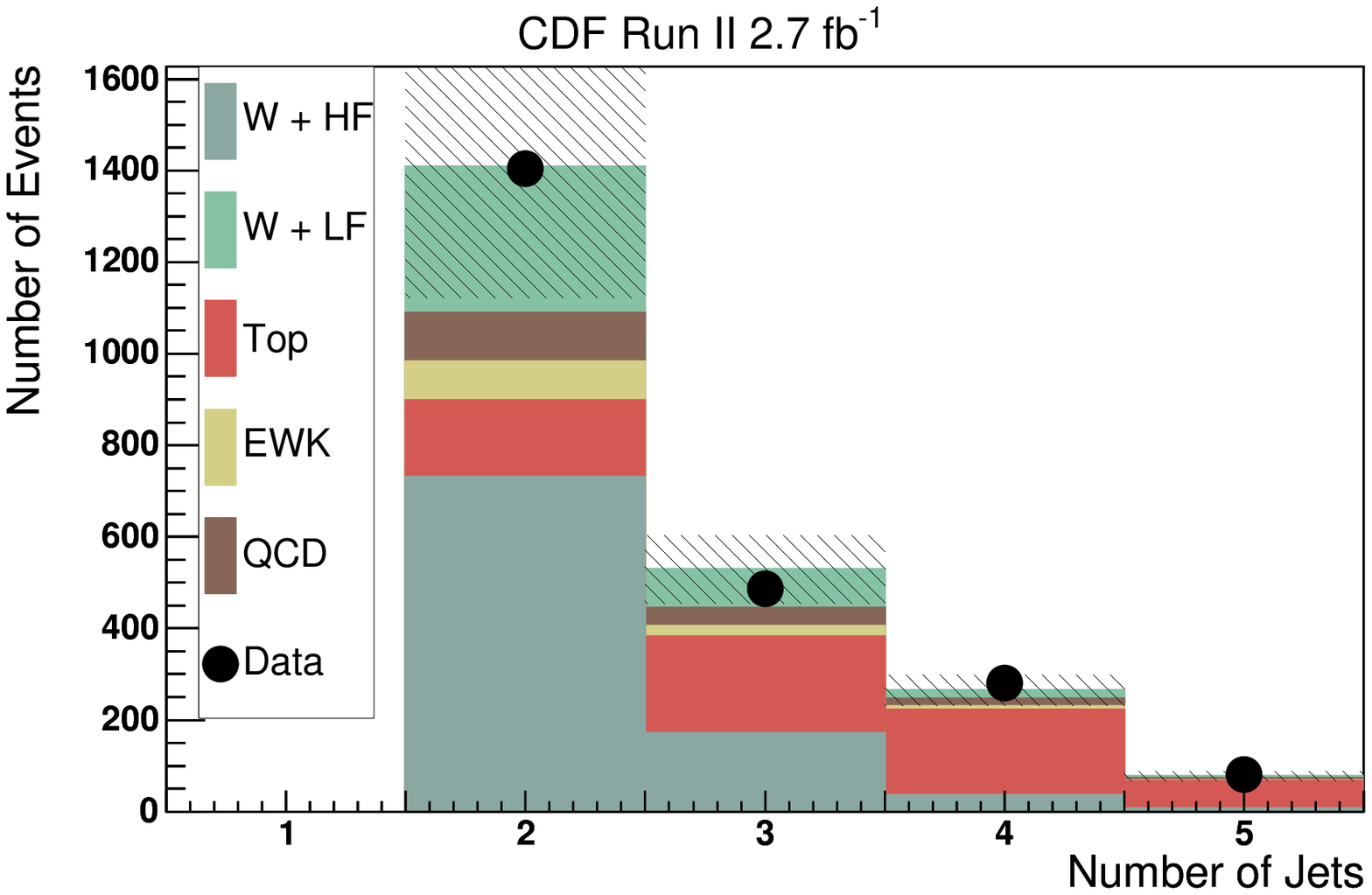}
    \includegraphics*[width=0.9\textwidth]{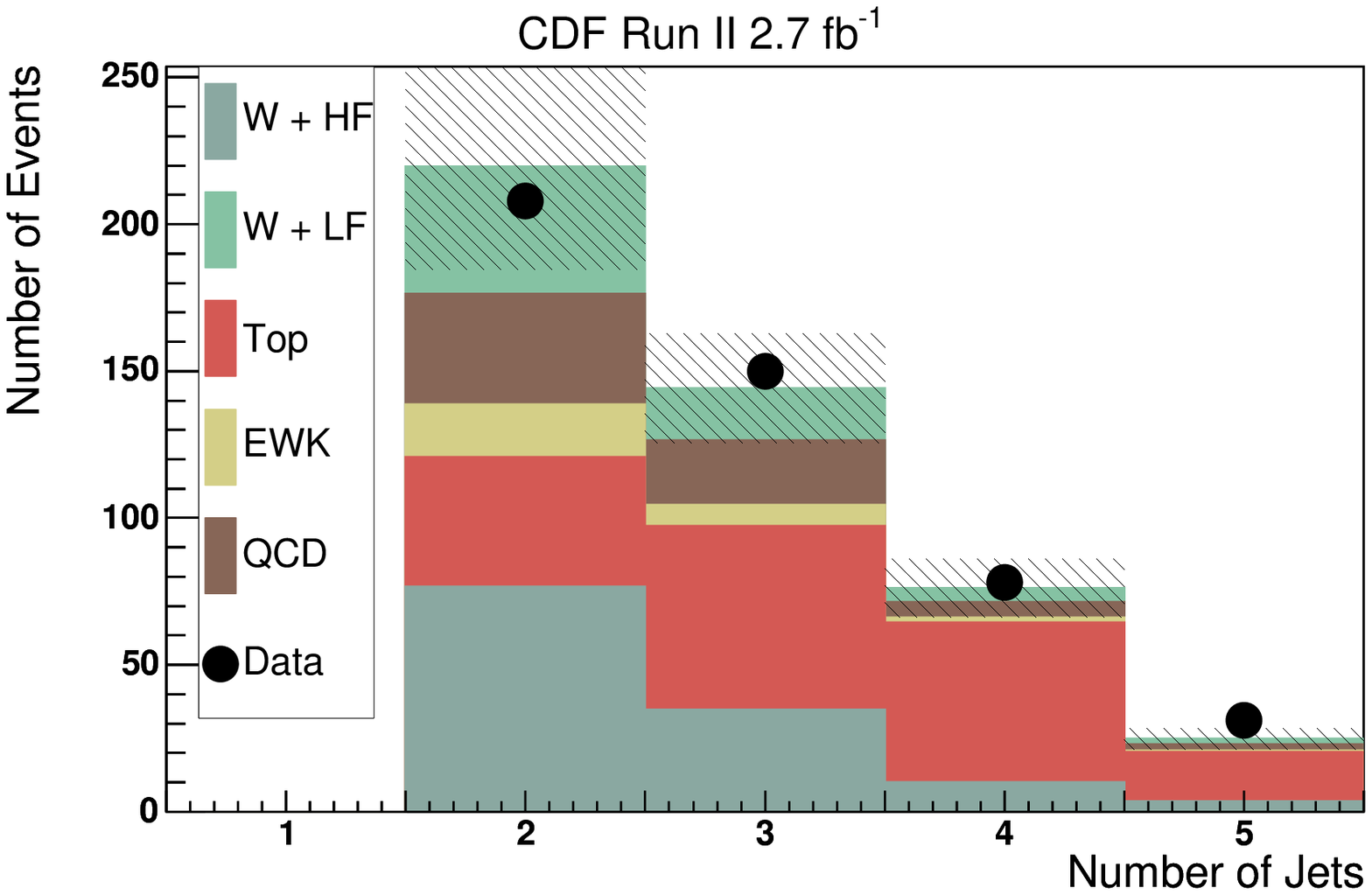}    
     \caption{Number of expected and observed background events for events
       with exactly one {\sc secvtx} tag, shown as a function of
       jet multiplicity. The plots show tight leptons (top)
       and isolated tracks (bottom). The hatched regions indicate the 
       total uncertainty.}
    \label{fig:Njets_eq1tag}
  \end{center}
\end{figure}

\begin{figure}[htbp]
  \begin{center}
     \includegraphics*[width=0.9\textwidth]{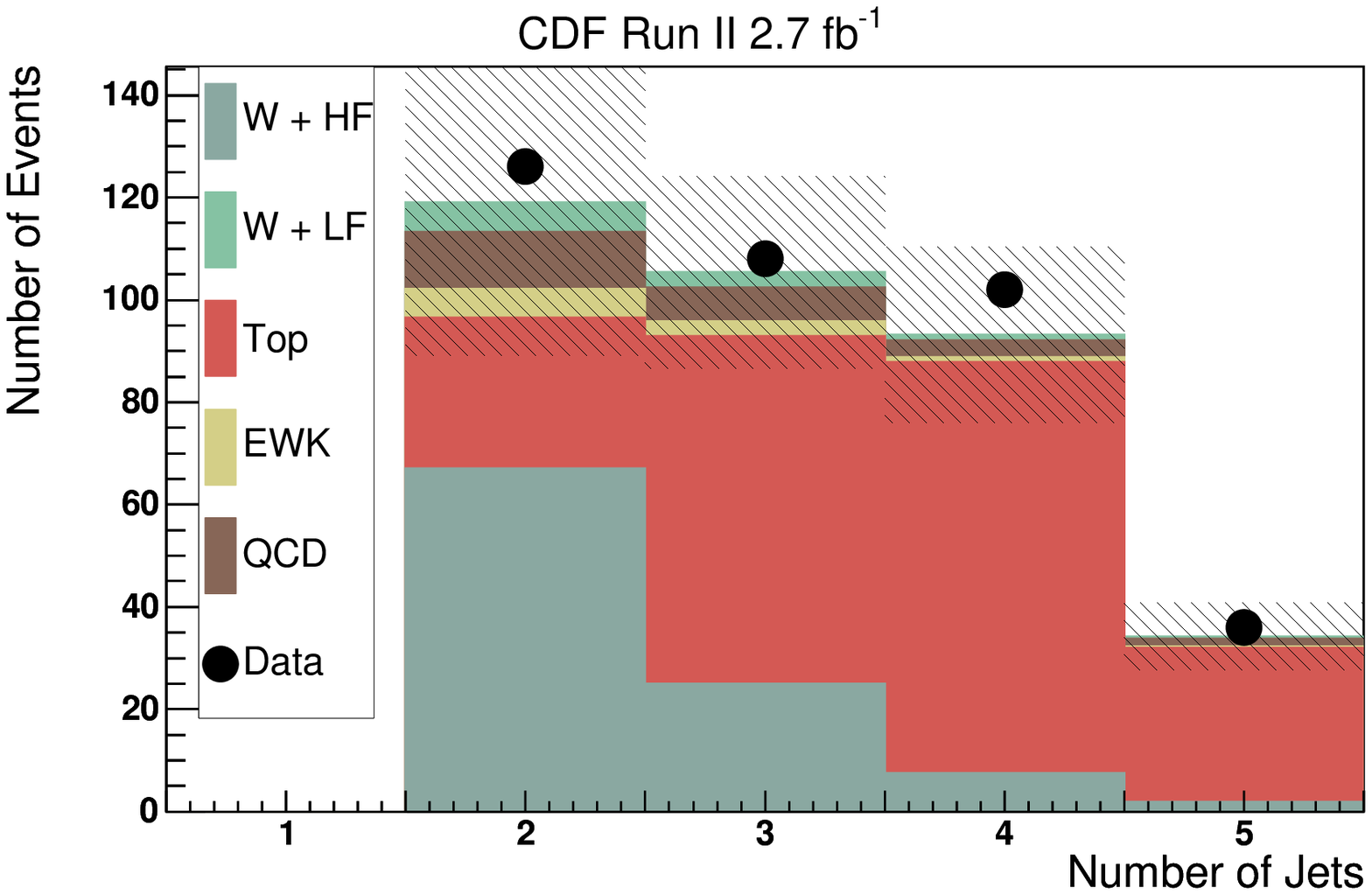}
     \includegraphics*[width=0.9\textwidth]{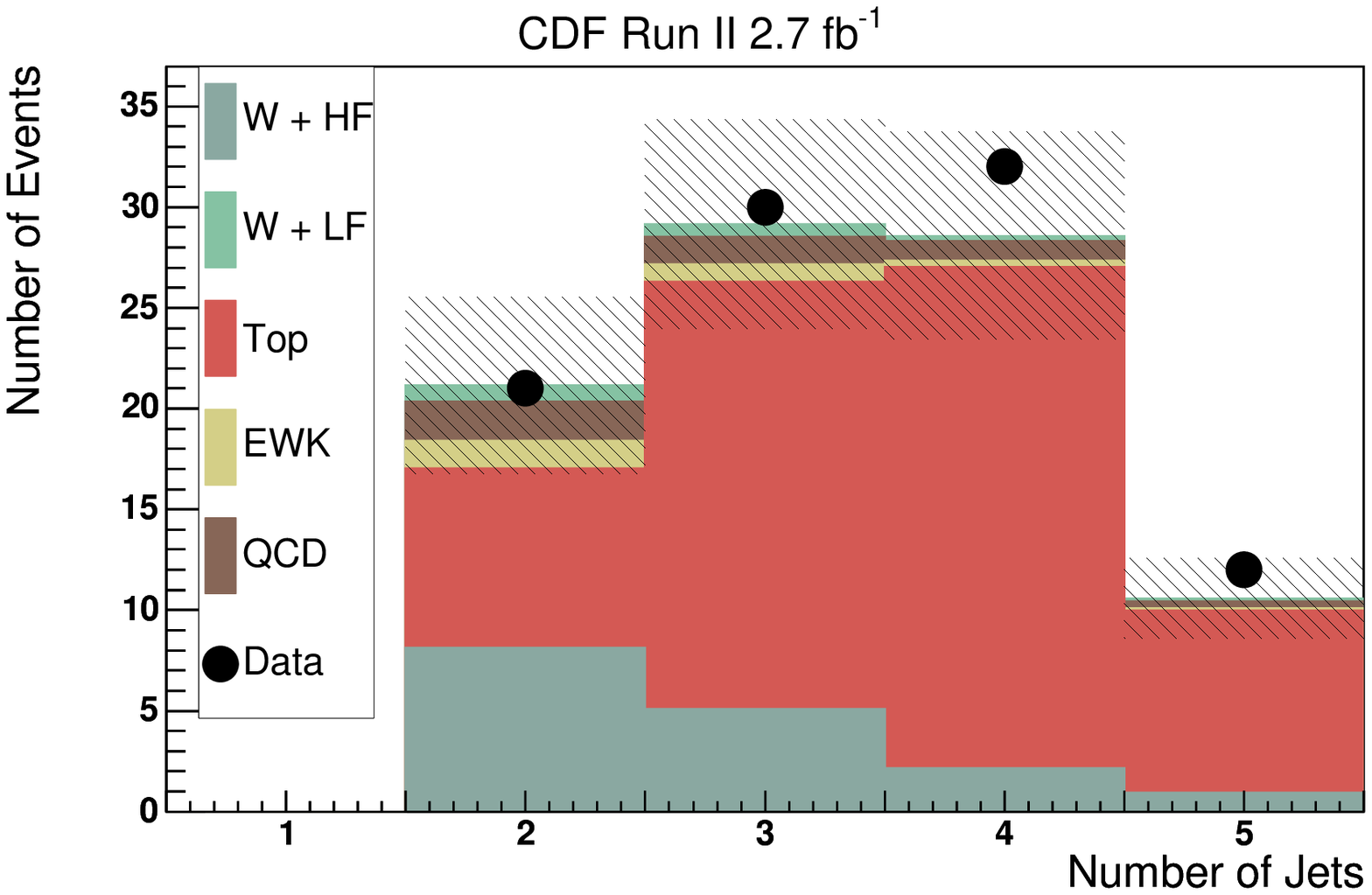}
    \caption{ Number of expected and observed background events for events
       with one {\sc secvtx} tag and one jetprob tag, shown as a function of
       jet multiplicity. The plots show tight leptons (top)
       and isolated tracks (bottom).The hatched regions indicate the 
       total uncertainty.}
    \label{fig:Njets_STJP}
  \end{center}
\end{figure}

\begin{figure}[htbp]
  \begin{center}
    \includegraphics*[width=0.9\textwidth]{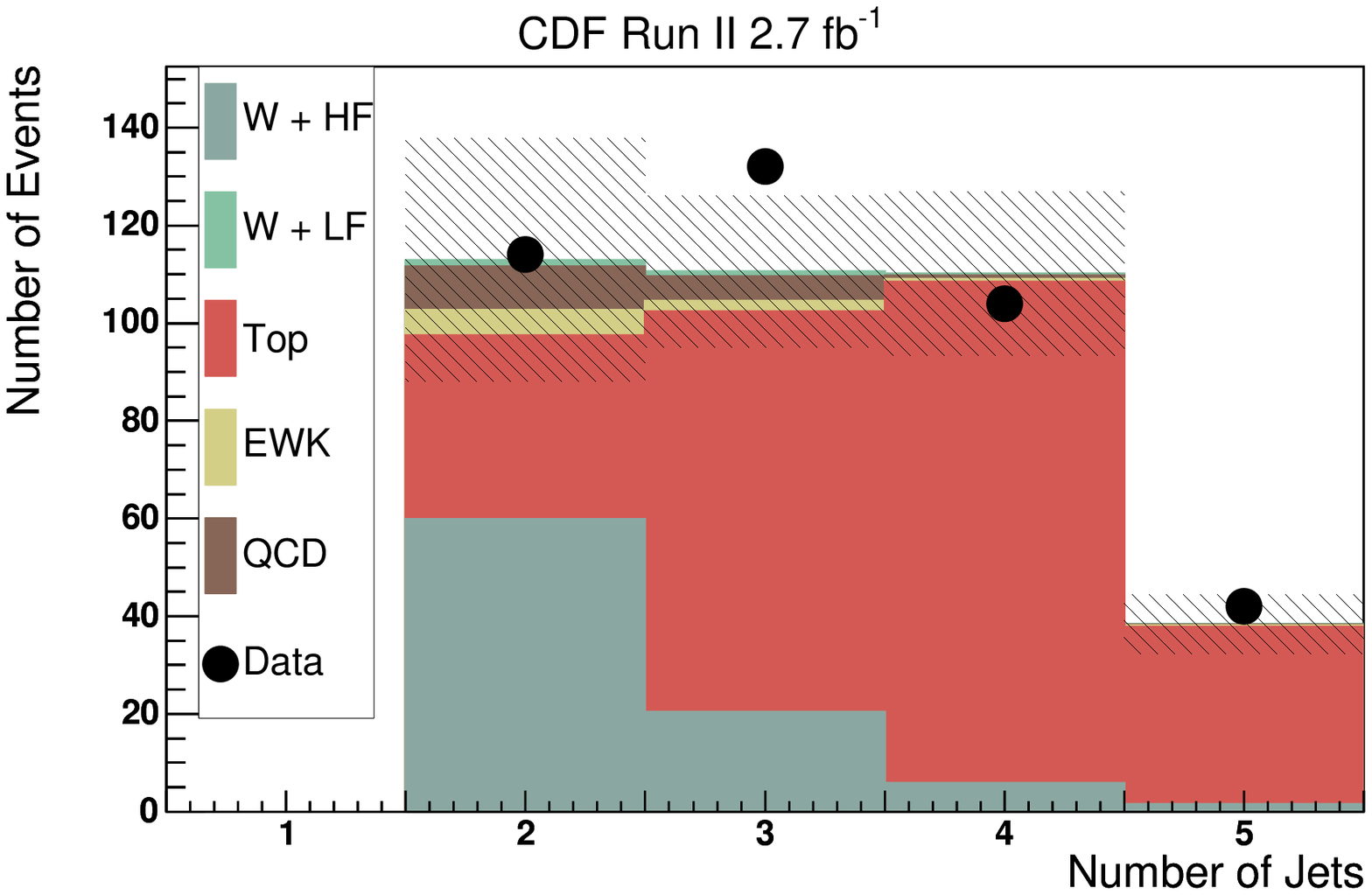}
    \includegraphics*[width=0.9\textwidth]{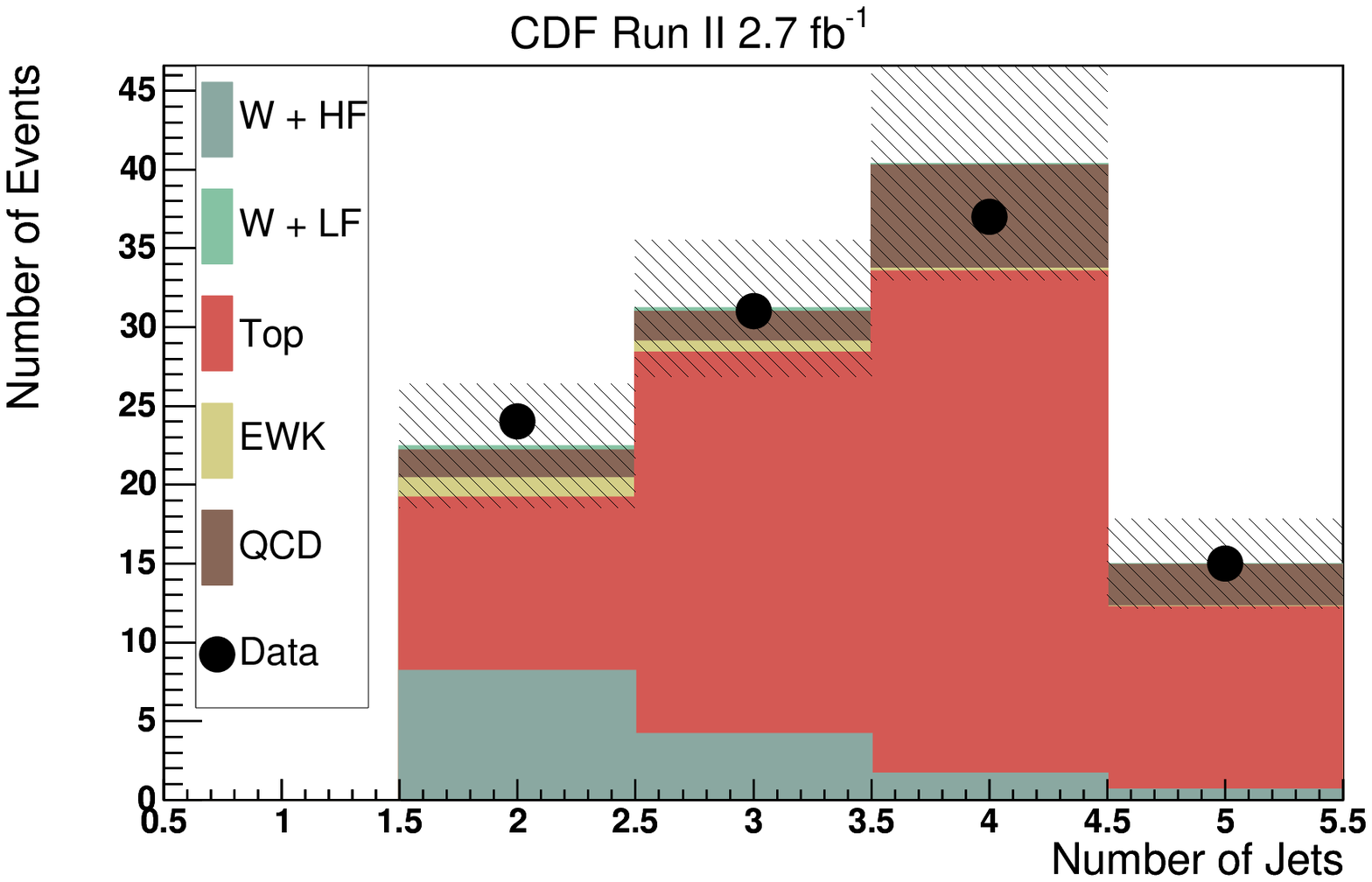}
    \caption{ Number of expected and observed background events for events
      with two {\sc secvtx} tags, shown as a function of
      jet multiplicity. The plots show tight leptons (top)
      and isolated tracks (bottom).The hatched regions indicate the 
       total uncertainty.}    
    \label{fig:Njets_STST}
  \end{center}
\end{figure}

Tables \ref{tbl:m2_tlep_eq1tag} through \ref{tbl:m2_isotrk_gr2tag}
summarize our background estimate for our dataset of 2.7
fb$^{-1}$. Figures \ref{fig:Njets_eq1tag} through \ref{fig:Njets_STST}
present the information from the tables as plots. The plots show the
background estimate compared to data. The largest errors on the
background estimate come from the large uncertainty on the heavy
flavor fraction used to calculate $W$ + charm and $W$ + bottom. We add
these large uncertainties linearly because they come from the same
source. The $b$-tagging scale factor uncertainty is also correlated
across all backgrounds and added linearly. In general, the background
estimate agrees with the data within uncertainties for each jet
multiplicity. The agreement of the background estimate with the data
in the high-jet-multiplicity bins gives us confidence that our
estimate is correct in our two-jet search region.

\begin{table}
\begin{center}
\begin{tabular}{ccccc}
\hline  
\hline
\multicolumn{5}{c}{CDF Run II  2.7 fb$^{-1}$}\\
\multicolumn{5}{c}{Tight Lepton Background Prediction and Event Yields}\\
\multicolumn{5}{c}{Events with Exactly One Secvtx Tag }\\
  \hline
Process & 2jets & 3jets & 4jets & 5jets \\
\hline 

          All Pretag Candidates &
          
          38729 &
          6380  &
          1677  &
          386

          \\

WW  & 40.6 $\pm$ 4.2 & 11.9 $\pm$ 1.2 & 2.92 $\pm$ 0.25 & 0.71 $\pm$ 0.06  \\ 
WZ  & 13.86 $\pm$ 0.94 & 3.43 $\pm$ 0.23 & 0.93 $\pm$ 0.06 & 0.2 $\pm$ 0.02  \\ 
ZZ  & 0.48 $\pm$ 0.04 & 0.19 $\pm$ 0.06 & 0.081 $\pm$ 0.007 & 0.023 $\pm$ 0.002  \\ 
Top Pair  & 102 $\pm$ 14 & 193 $\pm$ 26 & 183 $\pm$ 26 & 59.4 $\pm$ 8.8  \\ 
Single Top s-Channel  & 23.88 $\pm$ 2.2 & 6.95 $\pm$ 0.67 & 1.47 $\pm$ 0.15 & 0.28 $\pm$ 0.03  \\ 
Single Top t-Channel  & 42.53 $\pm$ 4.4 & 9.24 $\pm$ 0.94 & 1.62 $\pm$ 0.17 & 0.22 $\pm$ 0.02  \\ 
Z+Jets  & 28.72 $\pm$ 3.4 & 8.65 $\pm$ 0.96 & 2.73 $\pm$ 0.29 & 0.53 $\pm$ 0.06  \\ 
W+bottom  & 365.6 $\pm$ 140 & 91.0 $\pm$ 35 & 19.4 $\pm$ 8 & 3.97 $\pm$ 1.7  \\ 
W+charm  & 364.6 $\pm$ 140 & 81.2 $\pm$ 31 & 17.3 $\pm$ 7 & 3.64 $\pm$ 1.6  \\ 
Mistags  & 319 $\pm$ 42 & 83.8 $\pm$ 13 & 18.8 $\pm$ 5.07 & 3.82 $\pm$ 1.5  \\ 
Non-W  & 107 $\pm$ 43 & 40.2 $\pm$ 17 & 17.3 $\pm$ 14 & 4.48 $\pm$ 4.4  \\ 
\hline
Total Prediction  & 1408 $\pm$ 287 & 530 $\pm$ 75 & 266 $\pm$ 34 & 77 $\pm$ 11  \\

          Observed &
          
          1404 &
          486  &
          281  &
          81
          
          \\
          
\hline
\hline
  \end{tabular}
\end{center}
\caption{Background summary table for events with a central lepton 
and exactly one secondary vertex tag. The heavy-flavor fraction $F_{HF}$ 
is the source of the large correlated uncertainty for $W$+bottom and $W$+charm.
The other large source of correlated uncertainty is the $b$-tagging scale factor.
\label{tbl:m2_tlep_eq1tag}}
\end{table}

\begin{table}
\begin{center}
\begin{tabular}{ccccc}
\hline
\hline  
\multicolumn{5}{c}{CDF Run II  2.7 fb$^{-1}$}\\
\multicolumn{5}{c}{Isolated Track Background Prediction and Event Yields}\\
\multicolumn{5}{c}{Events with Exactly One Secvtx Tag }\\
  \hline
Process & 2jets & 3jets & 4jets & 5jets \\
\hline 

All Pretag Candidates &


4253 &
1380 &
427 &
117 

\\
WW  & 6.4 $\pm$ 0.65 & 2.83 $\pm$ 0.25 & 0.75 $\pm$ 0.07 & 0.23 $\pm$ 0.02  \\ 
WZ  & 2.41 $\pm$ 0.16 & 0.92 $\pm$ 0.06 & 0.19 $\pm$ 0.01 & 0.063 $\pm$ 0.005  \\ 
ZZ  & 0.127 $\pm$ 0.009 & 0.052 $\pm$ 0.004 & 0.007 $\pm$ 0.001 & 0.006 $\pm$ 0.001  \\ 
Top Pair  & 28.0 $\pm$ 3.8 & 58.3 $\pm$ 8.0 & 53.4 $\pm$ 7.6 & 16.8 $\pm$ 2.5  \\ 
Single Top s-Channel  & 6.08 $\pm$ 0.58 & 1.91 $\pm$ 0.19 & 0.43 $\pm$ 0.04 & 0.08 $\pm$ 0.01  \\ 
Single Top t-Channel  & 10.1 $\pm$ 1.1 & 2.32 $\pm$ 0.24 & 0.41 $\pm$ 0.05 & 0.07 $\pm$ 0.01  \\ 
Z+Jets  & 9.05 $\pm$ 1.1 & 3.35 $\pm$ 0.36 & 0.74 $\pm$ 0.077 & 0.16 $\pm$ 0.02  \\ 

W+bottom  & 39.9 $\pm$ 16 & 18.4 $\pm$ 7.3 & 5.35 $\pm$ 2.3 & 1.91 $\pm$ 0.79  \\ 
W+charm  & 36.7 $\pm$ 15 & 16.2 $\pm$ 6.5 & 4.66 $\pm$ 2.0 & 1.53 $\pm$ 0.64  \\ 
Mistags  & 43.2 $\pm$ 8.2 & 17.7 $\pm$ 4.0 & 4.81 $\pm$ 1.7 & 1.82 $\pm$ 0.64  \\ 
Non-W  & 37.6 $\pm$ 15 & 22.2 $\pm$ 8.9 & 5.26 $\pm$ 4.2 & 2.13 $\pm$ 1.7  \\ 
\hline
Total Prediction  & 220 $\pm$ 35 & 144 $\pm$ 19 & 76 $\pm$ 10 & 25 $\pm$ 3.4  \\

        Observed &
        208 &
        150 &
        78 &
        31 
        
          \\
\hline
\hline
  \end{tabular}
\end{center}
\caption{Background summary table for events with an isolated track and exactly  
one secondary vertex tag .
The heavy-flavor fraction $F_{HF}$ 
is the source of the large correlated uncertainty for $W$+bottom and $W$+charm.
The other large source of correlated uncertainty is the $b$-tagging scale factor.
\label{tbl:m2_isotrk_eq1tag}}
\end{table}

\begin{table}
\begin{center}
\begin{tabular}{ccccc}
\hline  
\hline
\multicolumn{5}{c}{CDF Run II  2.7 fb$^{-1}$}\\
\multicolumn{5}{c}{Tight Lepton Background Prediction and Event Yields}\\
\multicolumn{5}{c}{Events with One Secvtx Tag and  One Jet Prob Tag  }\\
  \hline
Process & 2jets & 3jets & 4jets & 5jets \\
\hline

          All Pretag Candidates &

          44723  &
          7573   &
          1677   &
          386  \\

 WW  & 1.24 $\pm$ 0.53 & 0.85 $\pm$ 0.31 & 0.4 $\pm$ 0.13 & 0.165 $\pm$ 0.047  \\ 
 WZ  & 2.51 $\pm$ 0.43 & 0.78 $\pm$ 0.16 & 0.18 $\pm$ 0.04 & 0.052 $\pm$ 0.013  \\ 
 ZZ  & 0.098 $\pm$ 0.017 & 0.053 $\pm$ 0.009 & 0.021 $\pm$ 0.004 & 0.005 $\pm$ 0.001  \\ 
 Top Pair  & 20.4 $\pm$ 4.2 & 63.9 $\pm$ 13 & 79.3 $\pm$ 16 & 29.9 $\pm$ 6.1  \\ 
 Single Top s-Channel  & 6.99 $\pm$ 1.1 & 2.45 $\pm$ 0.42 & 0.57 $\pm$ 0.1 & 0.133 $\pm$ 0.024  \\ 
 Single Top t-Channel  & 2.1 $\pm$ 0.64 & 1.67 $\pm$ 0.36 & 0.46 $\pm$ 0.09 & 0.076 $\pm$ 0.015  \\ 
 Z+Jets  & 1.81 $\pm$ 0.54 & 1.17 $\pm$ 0.35 & 0.34 $\pm$ 0.12 & 0.1 $\pm$ 0.03  \\ 

W+bottom  & 49.1 $\pm$ 20 & 17.1 $\pm$ 7.2 & 4.89 $\pm$ 2.1 & 1.28 $\pm$ 0.59  \\ 
W+charm  & 18.0 $\pm$ 8.3 & 7.89 $\pm$ 3.7 & 2.57 $\pm$ 1.2 & 0.67 $\pm$ 0.34  \\ 
Mistags  & 5.84 $\pm$ 6.0 & 3.01 $\pm$ 3.4 & 0.1 $\pm$ 1.1 & 0.29 $\pm$ 0.37  \\ 
Non-W  & 11.1 $\pm$ 5.33 & 6.57 $\pm$ 3.5 & 3.38 $\pm$ 3.4 & 1.51 $\pm$ 2.1  \\ 
\hline 
Total Prediction  & 119 $\pm$ 30 & 105 $\pm$ 19 & 93 $\pm$ 17 & 34 $\pm$ 7  \\

Observed & 124  & 109 & 101 & 36 \\

\hline
\hline
  \end{tabular}
\end{center}
\caption{Background summary table for events with a central lepton and two tags: one secondary vertex tag and one
  jet probability tag. The heavy-flavor fraction $F_{HF}$ 
is the source of the large correlated uncertainty for $W$+bottom and $W$+charm.
The other large source of correlated uncertainty is the $b$-tagging scale factor.}
\label{tbl:m2_tlep_stjptag}
\end{table}

\begin{table}
\begin{center}
\begin{tabular}{ccccc}
\hline
\hline  
\multicolumn{5}{c}{CDF Run II  2.7 fb$^{-1}$}\\
\multicolumn{5}{c}{Isolated Track Background Prediction and Event Yields}\\
\multicolumn{5}{c}{Events with One Secvtx Tag, One Jet Prob Tag  }\\
  \hline
Process & 2jets & 3jets & 4jets & 5jets \\
\hline

All Pretag Candidates &

5149 &
1623 &
487  &
124  

        \\

WW  & 0.2 $\pm$ 0.09 & 0.24 $\pm$ 0.09 & 0.1 $\pm$ 0.03 & 0.03 $\pm$ 0.01  \\ 
WZ  & 0.51 $\pm$ 0.09 & 0.2 $\pm$ 0.04 & 0.048 $\pm$ 0.01 & 0.013 $\pm$ 0.004  \\ 
ZZ  & 0.032 $\pm$ 0.006 & 0.021 $\pm$ 0.005 & 0.007 $\pm$ 0.001 & 0.002 $\pm$ 0.001  \\ 
Top Pair  & 6.44 $\pm$ 1.3 & 20.0 $\pm$ 4.2 & 24.6 $\pm$ 4.9 & 8.98 $\pm$ 1.8  \\ 
Single Top s-Channel  & 1.93 $\pm$ 0.31 & 0.74 $\pm$ 0.13 & 0.19 $\pm$ 0.03 & 0.043 $\pm$ 0.009  \\ 
Single Top t-Channel  & 0.53 $\pm$ 0.16 & 0.5 $\pm$ 0.11 & 0.12 $\pm$ 0.03 & 0.028 $\pm$ 0.005  \\ 
Z+Jets  & 0.61 $\pm$ 0.2 & 0.41 $\pm$ 0.13 & 0.13 $\pm$ 0.04 & 0.039 $\pm$ 0.013  \\ 

W+bottom  & 6.0 $\pm$ 2.7 & 3.4 $\pm$ 1.6 & 1.37 $\pm$ 0.67 & 0.59 $\pm$ 0.26  \\ 
W+charm  & 2.14 $\pm$ 1.07 & 1.64 $\pm$ 0.86 & 0.77 $\pm$ 0.41 & 0.34 $\pm$ 0.17  \\ 
 Mistags  & 0.8 $\pm$ 1.18 & 0.61 $\pm$ 0.84 & 0.27 $\pm$ 0.31 & 0.13 $\pm$ 0.17  \\ 
 Non-W  & 1.97 $\pm$ 0.79 & 1.38 $\pm$ 0.55 & 0.99 $\pm$ 0.79 & 0.37 $\pm$ 0.5  \\ 
\hline 
Total Prediction  & 21 $\pm$ 4 & 29 $\pm$ 5 & 29 $\pm$ 5 & 11 $\pm$ 2  \\

          Observed &

          21 &
          30 &
          32 &
          12 
          
        \\
\hline
\hline
  \end{tabular}
\end{center}
\caption{Background summary table for events with an 
isolated track and two tags: one secondary vertex tag
  and one jet probability tag.
The heavy-flavor fraction $F_{HF}$ 
is the source of the large correlated uncertainty for $W$+bottom and $W$+charm.
The other large source of correlated uncertainty is the $b$-tagging scale factor.
\label{tbl:m2_isotrk_stjptag}}
\end{table}

\begin{table}
\begin{center}
\begin{tabular}{ccccc}
\hline  
\hline
\multicolumn{5}{c}{CDF Run II  2.7 fb$^{-1}$}\\
\multicolumn{5}{c}{Tight Lepton Background Prediction and Event Yields}\\
\multicolumn{5}{c}{Events with Two Secvtx Tags  }\\
  \hline
Process & 2jets & 3jets & 4jets & 5jets \\
\hline

All Pretag Candidates &


44723  &
7573  &
1677  &
386

        \\

WW  & 0.3 $\pm$ 0.06 & 0.29 $\pm$ 0.05 & 0.17 $\pm$ 0.03 & 0.08 $\pm$ 0.01  \\ 
WZ  & 3.32 $\pm$ 0.37 & 0.94 $\pm$ 0.11 & 0.19 $\pm$ 0.02 & 0.04 $\pm$ 0.01  \\ 
ZZ  & 0.1 $\pm$ 0.01 & 0.073 $\pm$ 0.008 & 0.019 $\pm$ 0.002 & 0.005 $\pm$ 0.001  \\ 
Top Pair  & 25.9 $\pm$ 4.2 & 76.8 $\pm$ 12 & 101 $\pm$ 16 & 36.1 $\pm$ 5.9  \\ 
Single Top s-Channel  & 9.55 $\pm$ 1.2 & 3.25 $\pm$ 0.41 & 0.72 $\pm$ 0.09 & 0.15 $\pm$ 0.02  \\ 
Single Top t-Channel  & 2.15 $\pm$ 0.3 & 1.9 $\pm$ 0.26 & 0.53 $\pm$ 0.07 & 0.1 $\pm$ 0.01  \\ 
Z+Jets  & 1.42 $\pm$ 0.2 & 0.95 $\pm$ 0.13 & 0.26 $\pm$ 0.04 & 0.085 $\pm$ 0.013  \\ 
W+bottom  & 55.0 $\pm$ 22 & 18.1 $\pm$ 7.4 & 4.88 $\pm$ 2.0 & 1.24 $\pm$ 0.55  \\ 
W+charm  & 4.87 $\pm$ 2.0 & 2.35 $\pm$ 1 & 0.94 $\pm$ 0.4 & 0.25 $\pm$ 0.12  \\ 
 Mistags  & 1.38 $\pm$ 0.39 & 0.93 $\pm$ 0.3 & 0.34 $\pm$ 0.12 & 0.11 $\pm$ 0.05  \\ 
 Non-W  & 8.96 $\pm$ 4.0 & 5.02 $\pm$ 2.0 & 0.74 $\pm$ 1.6 & 0.23 $\pm$ 1.5  \\ 
\hline 
Total Prediction  & 113 $\pm$ 25 & 111 $\pm$ 16 & 110 $\pm$ 17 & 38 $\pm$ 6  \\

          Observed &
          

          114  &
          132  &
          104  &
          42

          \\
          
\hline
\hline
  \end{tabular}
\end{center}
\caption{Background summary table for events with a central lepton 
and  two secondary vertex tags.
The heavy-flavor fraction $F_{HF}$ 
is the source of the large correlated uncertainty for $W$+bottom and $W$+charm.
The other large source of correlated uncertainty is the $b$-tagging scale factor.
\label{tbl:m2_tlep_gr2tag}}
\end{table}

\begin{table}
\begin{center}
\begin{tabular}{ccccc}
\hline  
\hline
\multicolumn{5}{c}{CDF Run II  2.7 fb$^{-1}$}\\
\multicolumn{5}{c}{Isolated Track  Background Prediction and Event Yields}\\
\multicolumn{5}{c}{Events with Two Secvtx Tags  }\\
  \hline
Process & 2jets & 3jets & 4jets & 5jets \\
\hline 

All Pretag Candidates&


5149  &
1623  &
487   &
124 

        \\

WW  & 0.036 $\pm$ 0.008 & 0.13 $\pm$ 0.02 & 0.067 $\pm$ 0.012 & 0.019 $\pm$ 0.003  \\ 
WZ  & 0.65 $\pm$ 0.07 & 0.24 $\pm$ 0.03 & 0.029 $\pm$ 0.003 & 0.01 $\pm$ 0.001  \\ 
ZZ  & 0.045 $\pm$ 0.005 & 0.025 $\pm$ 0.003 & 0.01 $\pm$ 0.001 & 0.002 $\pm$ 0  \\ 
Top Pair  & 7.75 $\pm$ 1.2 & 22.7 $\pm$ 3.7 & 31.5 $\pm$ 5.1 & 11.5 $\pm$ 1.9  \\ 
Single Top s-Channel  & 2.66 $\pm$ 0.34 & 0.91 $\pm$ 0.12 & 0.21 $\pm$ 0.03 & 0.045 $\pm$ 0.006  \\ 
Single Top t-Channel  & 0.58 $\pm$ 0.08 & 0.57 $\pm$ 0.08 & 0.18 $\pm$ 0.02 & 0.035 $\pm$ 0.005  \\ 
Z+Jets  & 0.51 $\pm$ 0.07 & 0.32 $\pm$ 0.05 & 0.093 $\pm$ 0.014 & 0.025 $\pm$ 0.004  \\ 

W+bottom  & 7.51 $\pm$ 3.3 & 3.59 $\pm$ 1.63 & 1.41 $\pm$ 0.66 & 0.53 $\pm$ 0.23  \\ 
W+charm  & 0.68 $\pm$ 0.3 & 0.56 $\pm$ 0.26 & 0.26 $\pm$ 0.13 & 0.18 $\pm$ 0.05  \\ 
 Mistags  & 0.27 $\pm$ 0.13 & 0.2 $\pm$ 0.1 & 0.089 $\pm$ 0.05 & 0.052 $\pm$ 0.026  \\ 
 Non-W  & 1.78 $\pm$ 0.71 & 1.89 $\pm$ 0.76 & 6.53 $\pm$ 5.2 & 2.65 $\pm$ 2.1  \\ 
\hline 
Total Prediction  & 22 $\pm$ 4 & 31 $\pm$ 4 & 40 $\pm$ 7 & 15 $\pm$ 3  \\

          Observed&

          24 &
          31 &
          37 &
          15
          
        \\
\hline

  \end{tabular}
\end{center}
\caption{Background summary table for events with an isolated track and two secondary vertex tags.
The heavy-flavor fraction $F_{HF}$ 
is the source of the large correlated uncertainty for $W$+bottom and $W$+charm.
The other large source of correlated uncertainty is the $b$-tagging scale factor.
\label{tbl:m2_isotrk_gr2tag}}
\end{table}

\clearpage

\section{Higgs Boson Signal Acceptance}
\label{sec:Acceptance}

We simulated the $WH$ signal kinematics using the {\sc pythia} Monte
Carlo program \cite{Sjostrand:2000wi}.  We generated signal Monte
Carlo samples for Higgs masses between 100 and
$150\,\mathrm{GeV}/c^2$.  The number of expected $WH\rightarrow
\ell\nu b\bar b$ events, $N$, is given by:
\begin{equation}
N = \epsilon \cdot \int {\cal{L}} dt \cdot \sigma (p \bar{p}
 \rightarrow WH)\cdot {\cal B}(H \rightarrow b \bar{b}), \label{eq:ExpEvt}
 \label{eq:WHexpect}
\end{equation}
where $\epsilon$, $\int {\cal{L}} dt$, $\sigma(p \bar{p} \rightarrow
WH)$, and ${\cal B}(H\rightarrow b\bar{b})$ are the event detection
efficiency, integrated luminosity, production cross section, and
branching ratio, respectively.  The production cross section and
branching ratio are calculated to next-to-leading order (NLO)
precision~\cite{Djouadi:1997yw}. 

The total event detection efficiency is composed of several
efficiencies: the primary vertex reconstruction efficiency, the
trigger efficiency, the lepton identification efficiency, the
$b$-tagging efficiency, and the event selection
efficiency~\cite{WH2FB}. Each efficiency is calibrated to match
observations. 

\begin{figure}[htbp]
 \begin{center}
   \includegraphics[width=0.6\textwidth]{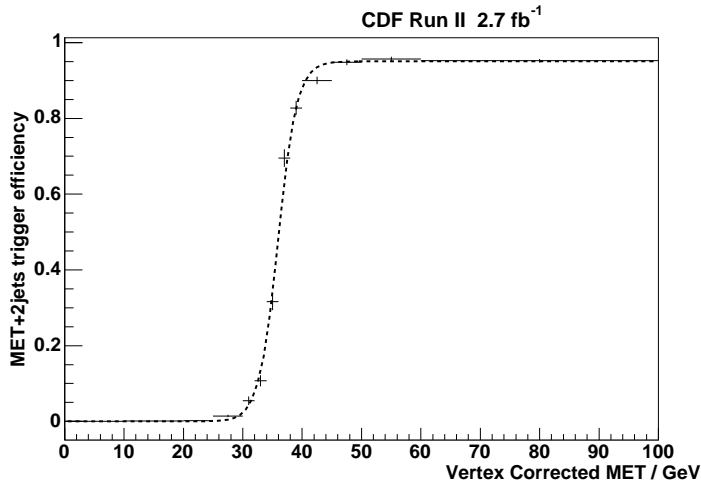}
   \caption{\MET plus jets trigger turn-on curve parameterized as a
     function of vertex \MET. The plot shows the turn-on curve measured
   in 2.7 fb$^{-1}$ of CDF data.}
   \label{fig:MetTrigTurnOn}
 \end{center}
\end{figure}

We parametrize the \MET trigger turn-on as a function of $\METvertex$,
which is \MET corrected for the primary vertex position but not muons
or jet energy scale corrections. We use $\METvertex$ because it is
close to the \MET calculation used by the trigger and is modeled
better in the Monte Carlo than $\METraw$, which is calculated assuming
$z_0 = 0$.  The measurement of the jets can influence the measurement
of the \MET.  We require that the jets in the event are above the
trigger threshold ($E_T > 25$~GeV) and well separated ($\Delta R >
1.0$), which reduces the impact of the jets on the \MET.  We measured
the turn-on curve using events recorded with the CMUP trigger, which
is independent from the \MET + 2 jets trigger. We selected events
passing our jet requirements, and measured their efficiency to pass
the \MET + 2 jets trigger as a function of $\METvertex$. Figure
\ref{fig:MetTrigTurnOn} shows the measured \MET + 2 jets trigger
turn-on. We use the parmeterized turn-on curve to weight each Monte
Carlo event according to its efficiency to pass the trigger.

%
%

The expected number of signal events is estimated by
equation~\ref{eq:WHexpect} at each Higgs boson mass point.
Table \ref{tab:acceptance} shows the number of
expected WH events for $M_H = 120$ GeV/c$^{2}$ in
2.7~fb$^{-1}$.

\begin{table}
\begin{center}
  \begin{tabular}{cc}
    \hline
    \hline  
    \multicolumn{2}{c}{CDF Run II 2.7 fb$^{-1}$}\\
    \multicolumn{2}{c}{Number of Expected WH ($M_H$ = 120 GeV/c$^2$) Events}\\
    \hline 
     Lepton Type  & Expected Number of WH events \\
    \hline
    \multicolumn{2}{c}{Exactly One Secvtx Tag}\\
    \hline 
    CEM & 1.58 $\pm$  0.08\\
    CMUP & 0.91 $\pm$  0.05 \\
    CMX & 0.44 $\pm$ 0.02 \\
    ISOTRK & 0.72 $\pm$ 0.07 \\
    Total & 3.65 $\pm$  0.22 \\
    \hline
    \multicolumn{2}{c}{Two Secvtx Tags}\\
    \hline
    CEM & 0.66 $\pm$ 0.07\\
    CMUP & 0.37 $\pm$ 0.04 \\
    CMX & 0.17 $\pm$ 0.02\\
    ISOTRK & 0.36 $\pm$ 0.05\\    
    Total & 1.56 $\pm$ 0.18  \\
    \hline
    \multicolumn{2}{c}{One Secvtx Tag and One Jet Probability Tag}\\
    \hline
    CEM & 0.48 $\pm$ 0.05  \\
    CMUP & 0.26 $\pm$ 0.03   \\
    CMX & 0.13 $\pm$ 0.01  \\
    ISOTRK & 0.23 $\pm$  0.03 \\    
    Total & 1.10 $\pm$ 0.12 \\
    \hline
    \hline
  \end{tabular}
  \caption{Expected number of WH events at a M(H)=120,
  shown separately for different tag categories and lepton types. The 
  lepton types are categorized based on the sub-detector regions.}
  \label{tab:acceptance}
\end{center}
\end{table}

The total systematic uncertainty on the acceptance comes from several
sources, including the jet energy scale, initial and final state
radiation, lepton identification, trigger efficiencies, and
$b$-tagging scale factor. The largest uncertainties come from the
$b$-tagging scale factor uncertainty and isolated track identification
uncertainty.

We assign a 2\% uncertainty to the CEM, CMUP, and CMX lepton
identification efficiency, and an 8\% uncertainty to isolated track
identification.  The identification uncertainties are based on studies
comparing $Z$ boson events in data and Monte Carlo.

The high $p_T$ lepton triggers have a 1\% uncertainty on their
efficiencies.  We measure the trigger efficiency uncertainty by using
backup trigger paths or $Z$ boson events. We measure a 3\% uncertainty
for events collected on the \MET + 2 jets trigger by examining the
variations in the \MET turn-on curve in sub-samples with kinematics
different from the average sample. We use the variation in the \MET
turn-on to calculate a variation in signal acceptance, and we use the
mean variation in signal acceptance as our uncertainty.

We estimate the impact of changes in initial and final state radiation
by halving and doubling the parameters related to ISR and FSR in the
Monte Carlo event generation ~\cite{Abulencia:2005aj}.  The difference
from the nominal acceptance is taken as the systematic uncertainty.

The uncertainty in the incoming partons' energies relies on the the
parton distribution function (PDF) fits.  A NLO version of the PDFs,
CTEQ6M, provides a 90\% confidence interval of each
eigenvector~\cite{Pumplin:2002vw}. The nominal PDF value is reweighted
to the 90\% confidence level value, and the corresponding reweighted
acceptance is computed.  The differences between the nominal and the
reweighted acceptances are added in quadrature, and the total is
assigned as the systematic uncertainty~\cite{Acosta:2004hw}.

The uncertainty due to the jet energy scale uncertainty
(JES)~\cite{Bhatti:2005ai} is calculated by shifting jet energies in
$WH$ Monte Carlo samples by $\pm 1\sigma$.  The deviation from the
nominal acceptance is taken as the systematic uncertainty.

The systematic uncertainty on the $b$-tagging efficiency is based on
the scale factor uncertainty discussed in Sec.~\ref{sec:wPlusHeavy}.
The total systematic uncertainties for various $b$-tagging options and
lepton categories are summarized in Table~\ref{tbl:sys2jet}.

\begin{table}
\begin{center}
\begin{tabular}{ccccc} 
\hline
\hline
Source                      & \multicolumn{3}{c}{Uncertainty (\%)} \\ 
\hline
                            & Two Secvtx Tags   & One Secvtx One JetProb tag & Exactly One Secvtx   \\ \hline
Trigger Lepton (Isotrk) ID  & $\sim$2\% (8.85\%) & $\sim$2\% (8.85\%) & $\sim$2\% (8.85\%)  \\ 
Lepton (MET+Jets) Trigger   & $<$1\% (3\%) &    $<$1\% (3\%) & $<$1\% (3\%)   \\
ISR/FSR                     &        5.2\% &     4.0\% & 2.9\%     \\
PDF                         &        2.1\% &     1.5\% & 2.3\%     \\
JES                         &        2.5\% &     2.8\% & 1.2\%     \\
b-tagging                   &        8.4\% &     9.1\% & 3.5\%     \\
\hline
Total (Isotrk)    &        10.6\% (13.8\%) &    10.5\% (14.0\%) & 5.6\% (10.1\%) \\  
\hline
\hline
\end{tabular}
\end{center}
\caption{Systematic uncertainty on the $WH$ acceptance.  ``ST+ST''
refers to double secondary vertex tagged events while ``ST+JP'' refers
to secondary vertex plus jet probability tagged events.  Effects of
limited Monte Carlo statistics are included in these values.}
\label{tbl:sys2jet}
\end{table}


\section{Neural Network Discriminant}
\label{sec:neuralnet}
To further improve the signal to background discrimination after event
selection, we employ an artificial neural network (NN). Neural networks
offer an advantage over a single-variable discriminants because they
combine information from several kinematic variables. Our neural 
network is trained to distinguish $W$+Higgs boson events
from backgrounds.  We employ the same neural network that was used to
obtain the 1.9 fb$^{-1}$ result \cite{WH2FB}. The following section reviews
its main features.




Our neural network configuration has 6 input variables, 11 hidden nodes,
and 1 output node. The input variables were selected by an iterative 
network optimization procedure from a list of 76 possible variables. 
The optimization procedure identified the most sensitive one-variable 
NN, then looped over all remaining variables and found the most sensitive 
two-variable NN. The process continued until adding a new variable
does not improve sensitivity by more than 0.5 percent. The 6  inputs are:

\begin{description}
\item[$M_{jj+}$:] The dijet mass plus is the invariant mass calculated
  from the two reconstructed jets.  If there are additional loose jets
  present, where loose jets have $E_{T} > 12$ GeV, $|\eta| < 2.4$ and
  have a centroid within $\Delta R < 0.9$ of one of the leading jets,
  then the loose jet that is closest to one of the two jets is
  included in this invariant mass calculation. 
  \item[$\sum{E_T}$(Loose Jets):] This variable is the scalar sum of
  the loose jet transverse energies.
  \item[$p_T$ Imbalance:] This variable expresses the difference
  between \MET and the scalar sum of the transverse momenta of the lepton
  and the jets. Specifically, it is calculated as $
  P_T(jet_{1}) + P_T(jet_{2}) + P_T(lep) -$ \MET.
  \item[$M_{l\nu j}^{min}$:] This is the invariant mass of the lepton,
  \MET, and one of the two jets, where the jet is chosen to give the
  minimum invariant mass.  For this quantity, the $p_z$ component of the
  neutrino is ignored.
\item[$\Delta R$(lepton-$\nu_{max}$):] This is the $\Delta R$
  separation between the lepton and the neutrino. We calculate the
  $p_z$ of the neutrino by constraining the lepton and the \MET to the
  $W$ mass (80.42 GeV/$c^{2}$). The constraint produces a quadratic
  equation for $p_Z$ and we choose the larger solution.
  \item[$P_T(W+H)$:] This is the total transverse momentum of the $W$
  plus two jets system, $P_T(\vec{lep} + \vec{\nu} + \vec{jet_{1}} +
  \vec{jet_{2}})$.
\end{description}
The strongest discriminating variable in the neural network is the
dijet mass plus.


We train our neural network with $W$+jets, $t\bar{t}$, single top, and
$WH$ signal Monte Carlo. We do not use QCD events to train our neural
network. We use the same topology and input variables to train
separate neural networks for each Higgs signal Monte Carlo sample. The
samples range from $M(H) =$ 100 to 150 GeV/c$^2$ in 5 GeV
increments. At each Higgs mass, we use the same neural network for
tight lepton and isolated track events.

\begin{figure}[htbp]
  \begin{center}
    \includegraphics*[width=0.48\textwidth]{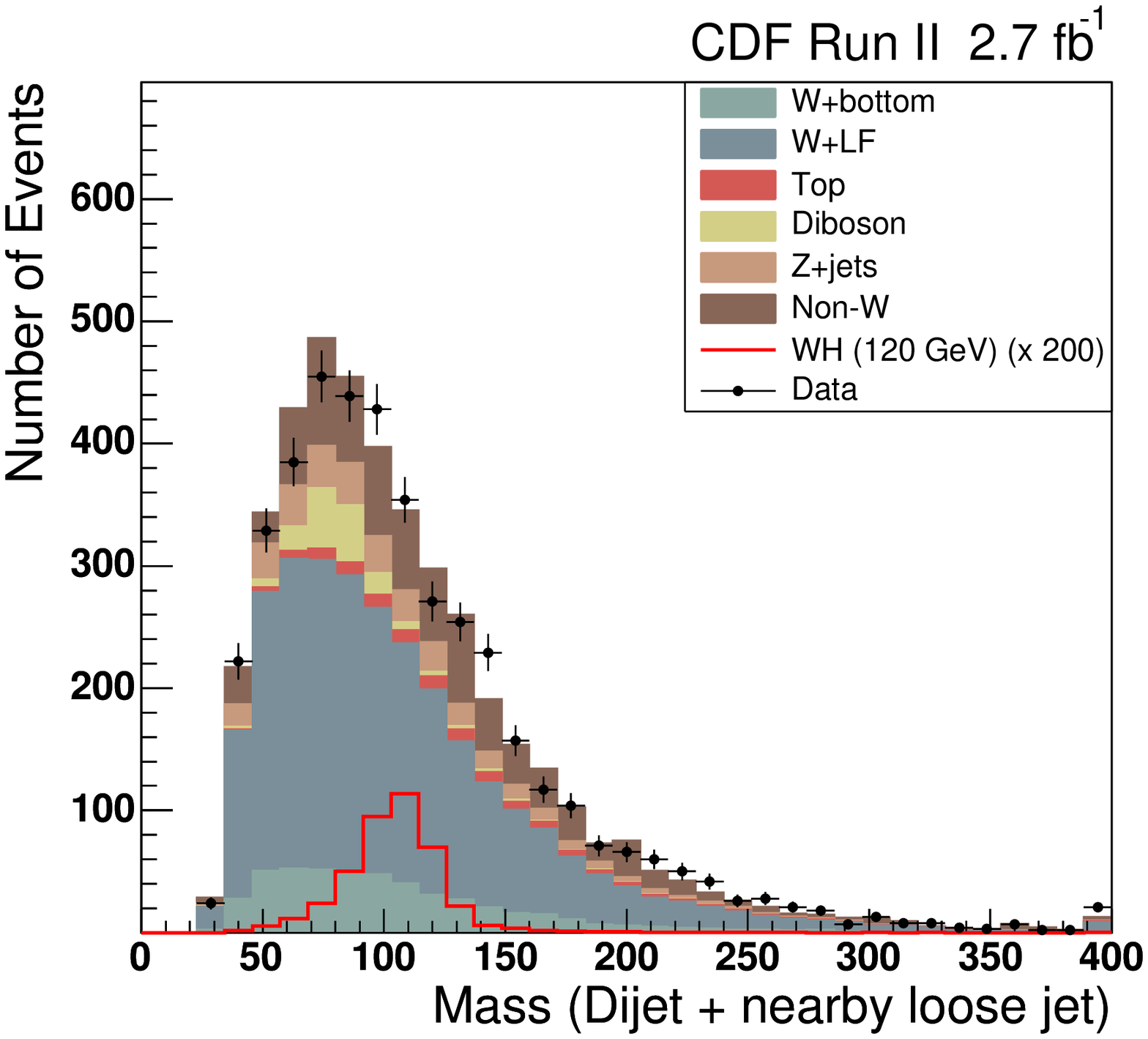}
    \includegraphics*[width=0.48\textwidth]{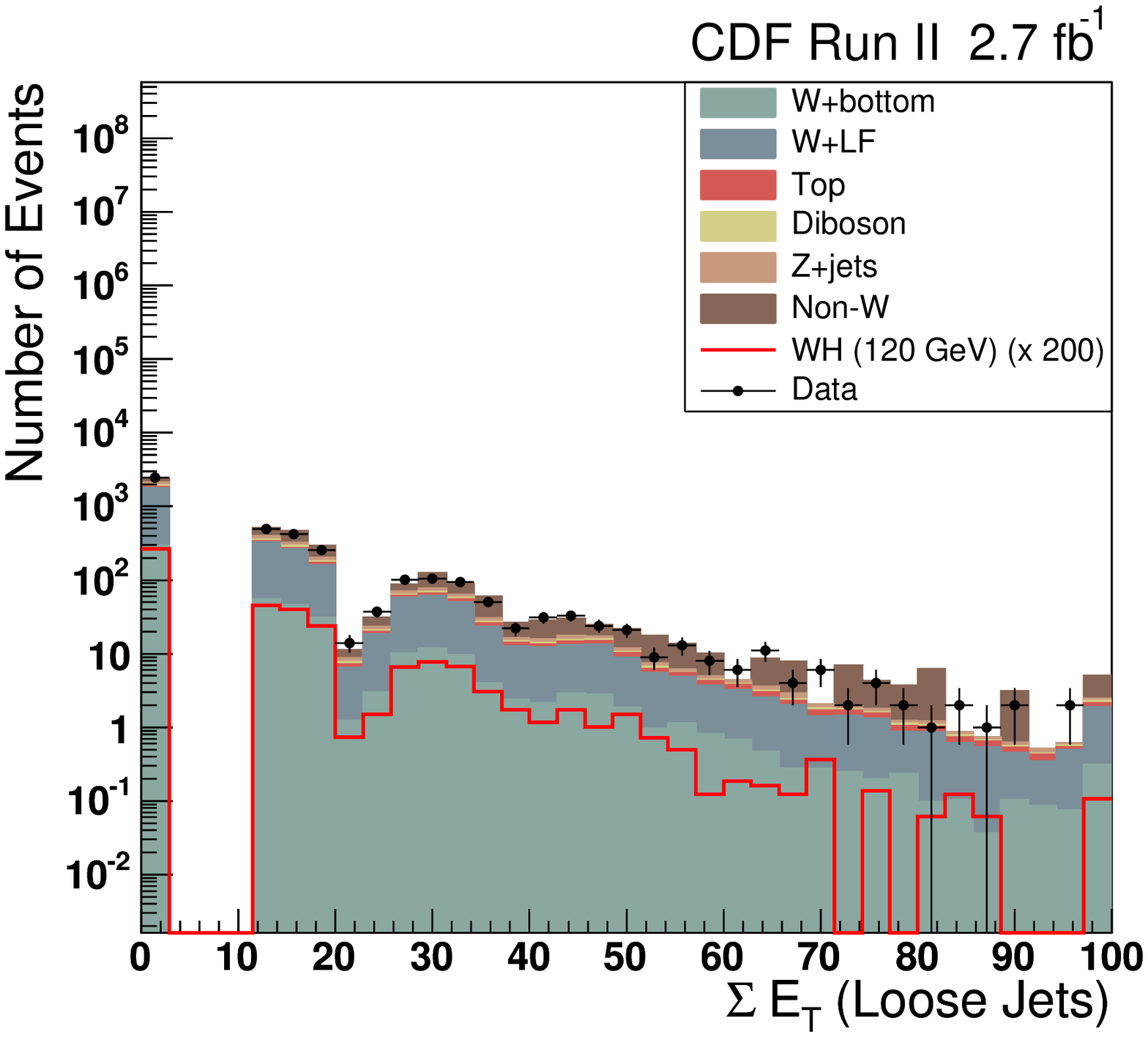}

     \caption{ Neural network input distributions for isolated track
     $W+2$ jet events in the pretag control region. The distributions
     shown are $M_{jj+}$ (left) and  $\sum{E_T}$(Loose Jets)
     (right). The differences in shape are attributable to QCD and 
     are less significant in our higher-purity search regions. }
    \label{fig:NNinput1}
  \end{center}
\end{figure}

\begin{figure}[htbp]
  \begin{center}
    \includegraphics*[width=0.48\textwidth]{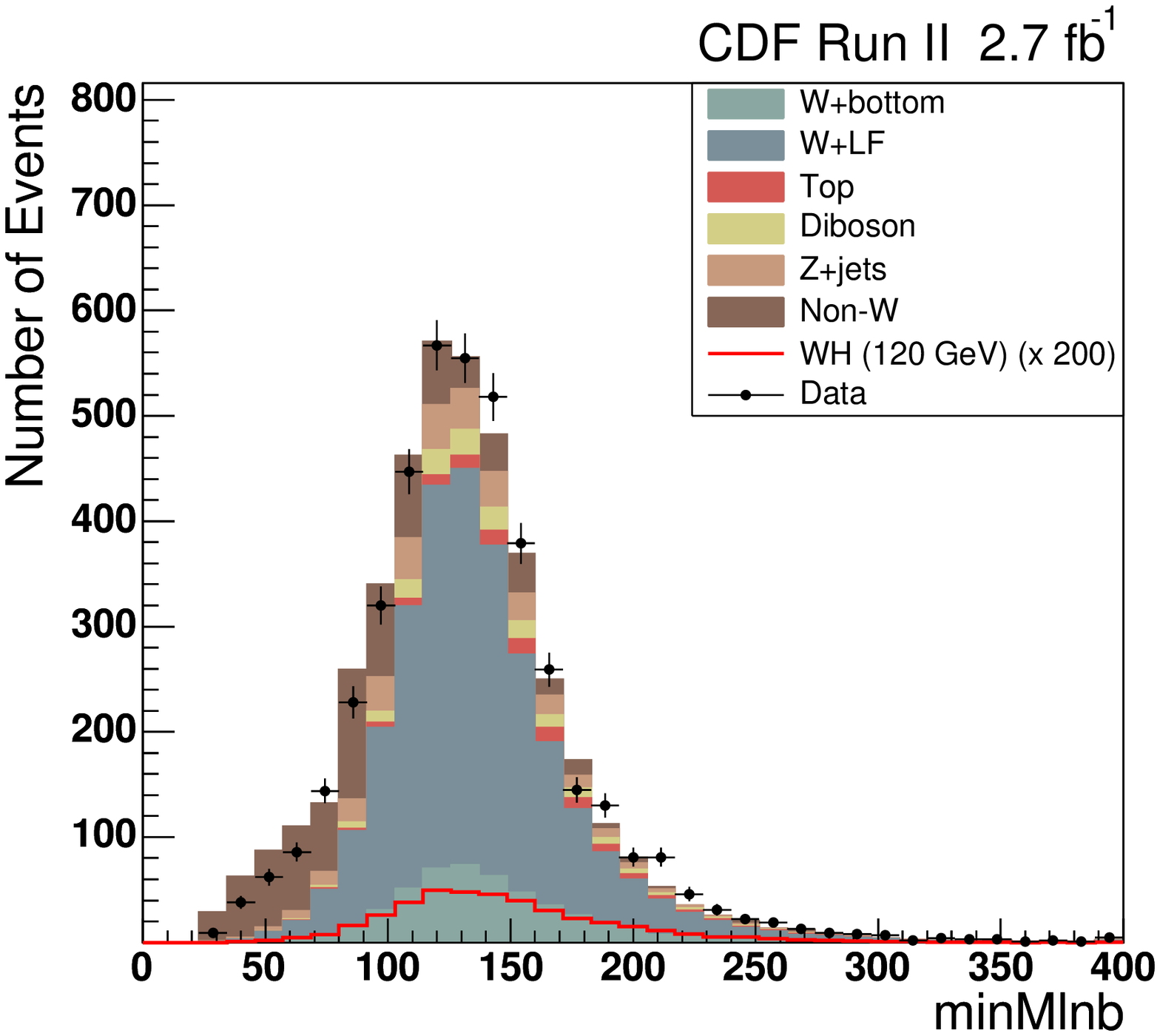}
    \includegraphics*[width=0.48\textwidth]{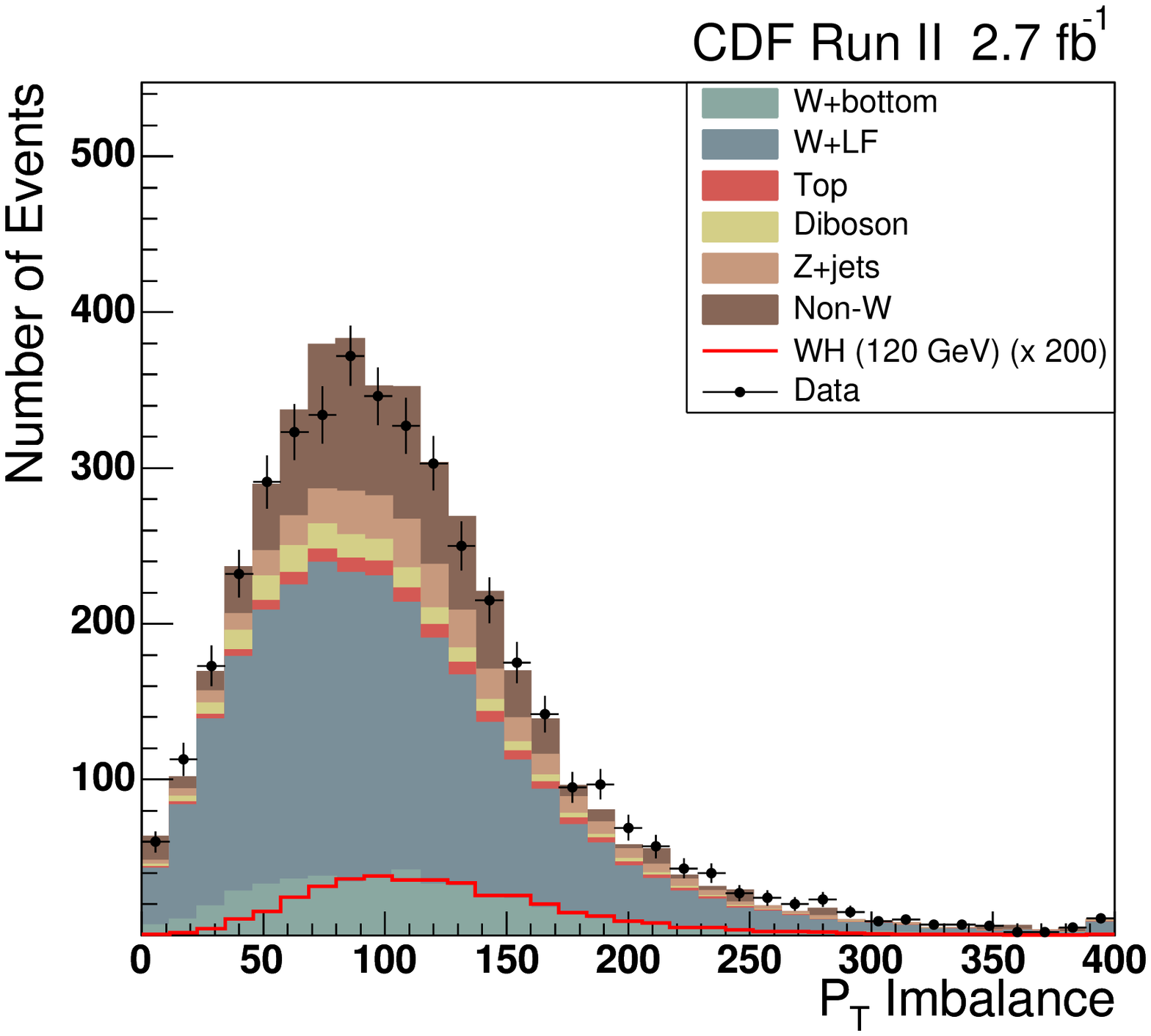}
    \caption{ Neural network input distributions for isolated track
      $W+2$ jet events in the pretag control region. The distributions
      shown are $M_{l\nu j}^{min}$ (left) and $P_T$ Imbalance (right).
      The differences in shape are attributable to QCD and 
      are less significant in our higher-purity search regions.
      \label{fig:NNinput2}
    }

  \end{center}
\end{figure}

\begin{figure}[htbp]
  \begin{center}

    \includegraphics*[width=0.48\textwidth]{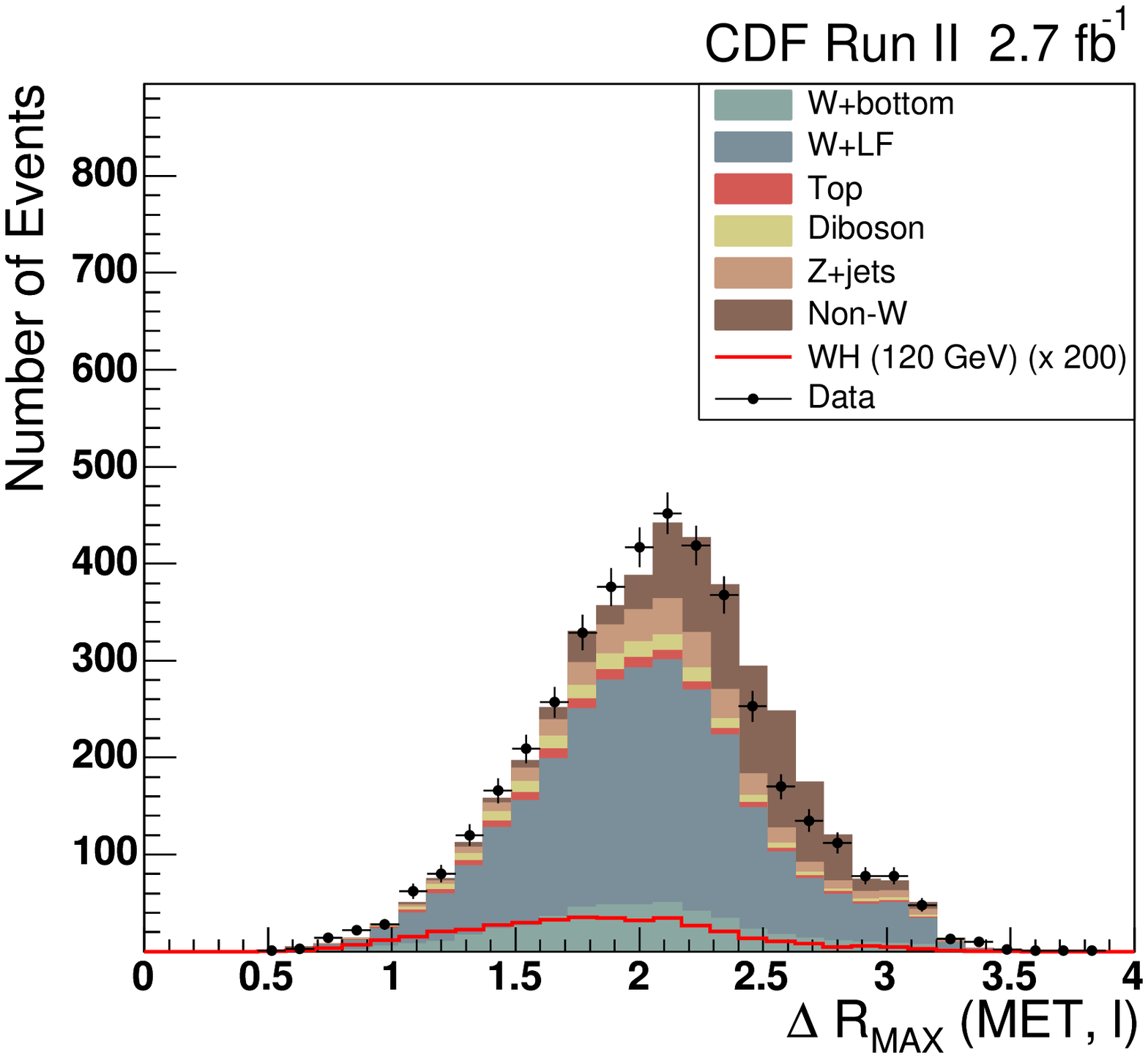} 
    \includegraphics*[width=0.48\textwidth]{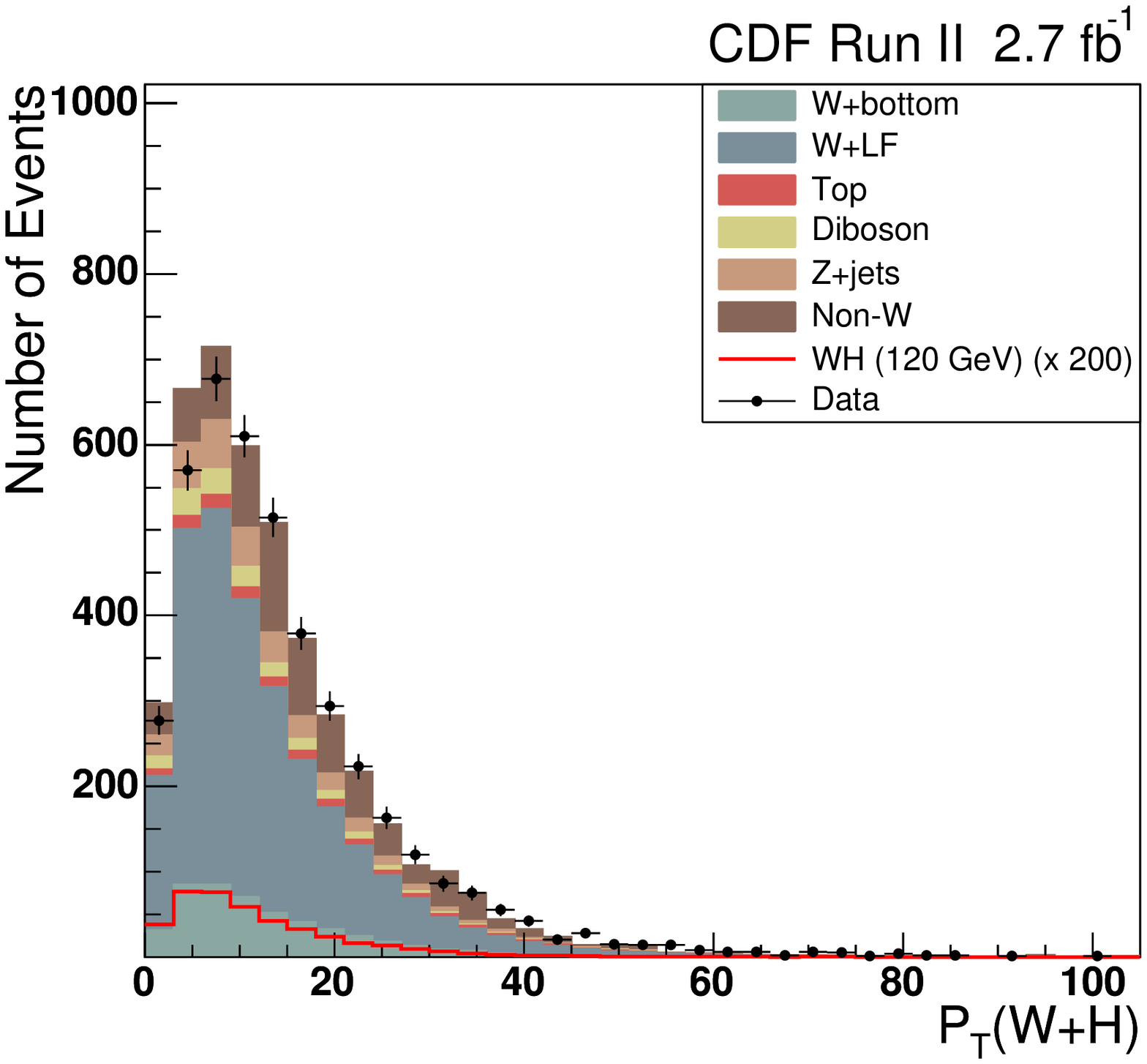}
    \caption{Neural network input distributions for isolated track
      $W+2$ jet events in the pretag control region. The distributions
      shown are $\Delta R$(lepton-$\nu_{max}$) (left), $P_T(W+H)$
      (right). The differences in shape are attributable to QCD and 
      are less significant in our higher-purity search regions.
      \label{fig:NNinput3}
  }

  \end{center}
\end{figure}

\begin{figure}[htbp]
  \begin{center}
    \includegraphics*[width=0.48\textwidth]{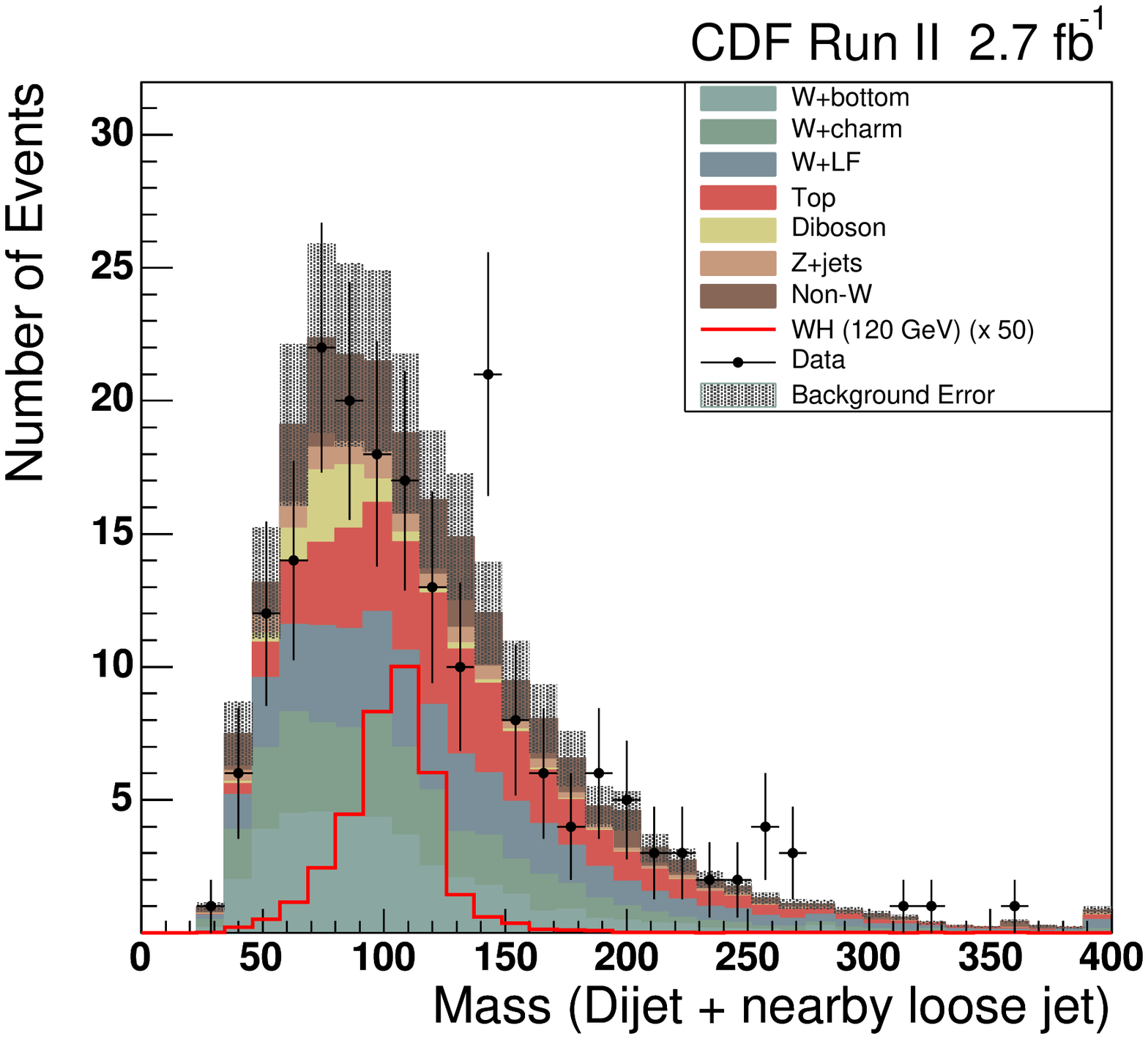}
    \includegraphics*[width=0.48\textwidth]{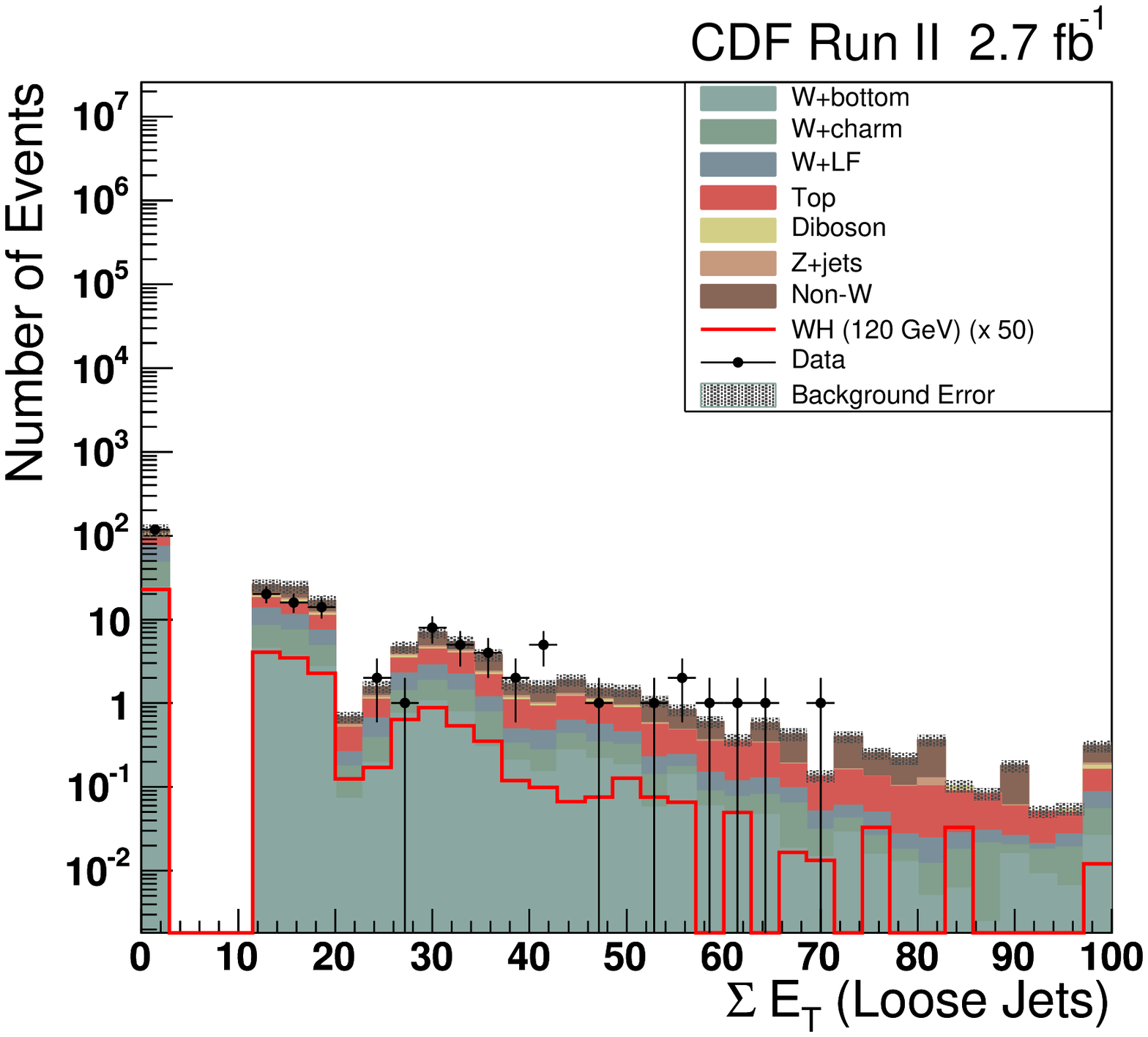}

     \caption{ Neural network input distributions for isolated track
     $W+2$ jet events in the one SECVTX tag region. The distributions
     shown are $M_{jj+}$ (left) and  $\sum{E_T}$(Loose Jets)
     (right). The differences in the shape are consistent with 
     the uncertainty on our QCD model.
     \label{fig:NNinput4}
   }
    
  \end{center}
\end{figure}

\begin{figure}[htbp]
  \begin{center}
    \includegraphics*[width=0.48\textwidth]{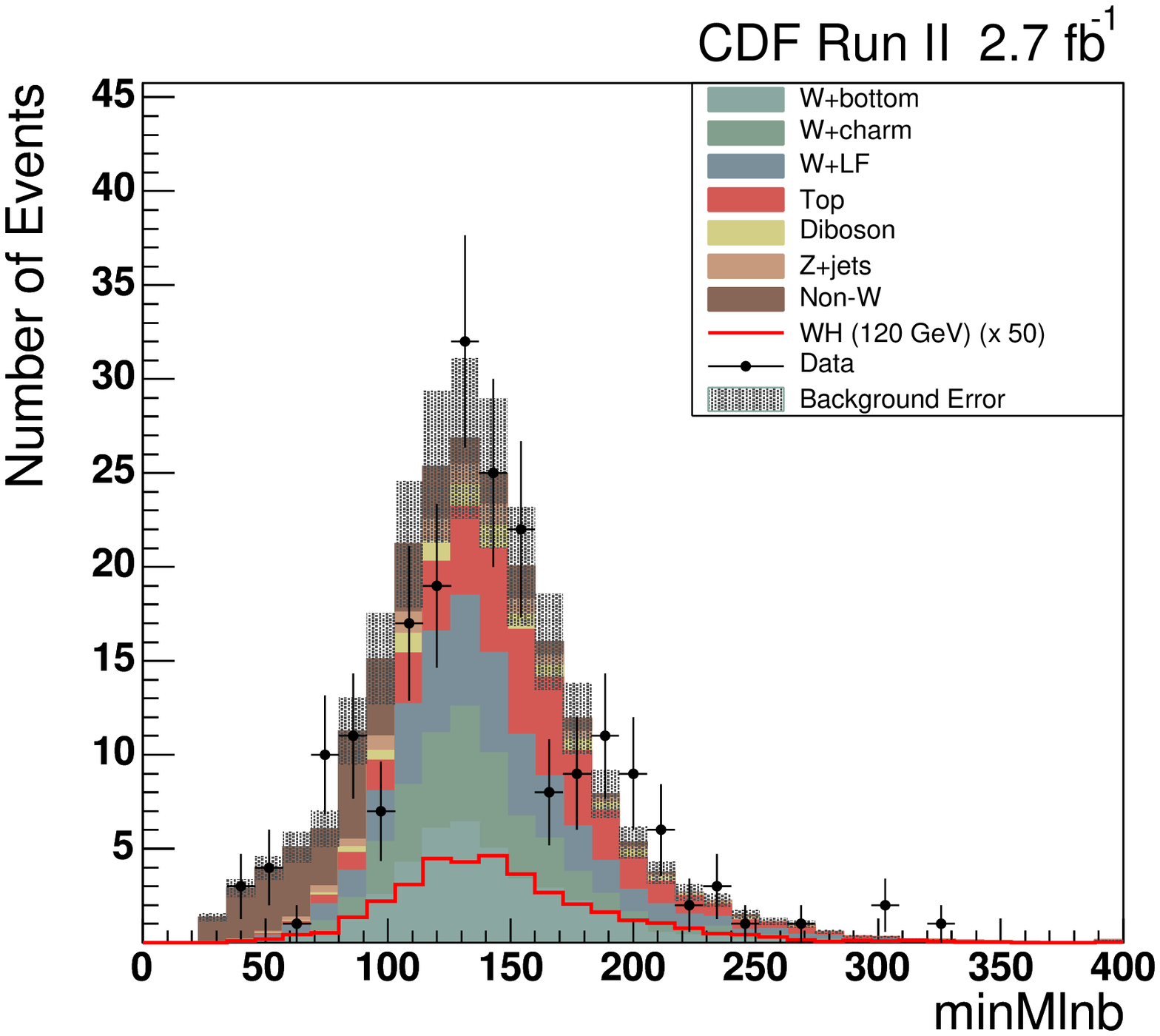}
    \includegraphics*[width=0.48\textwidth]{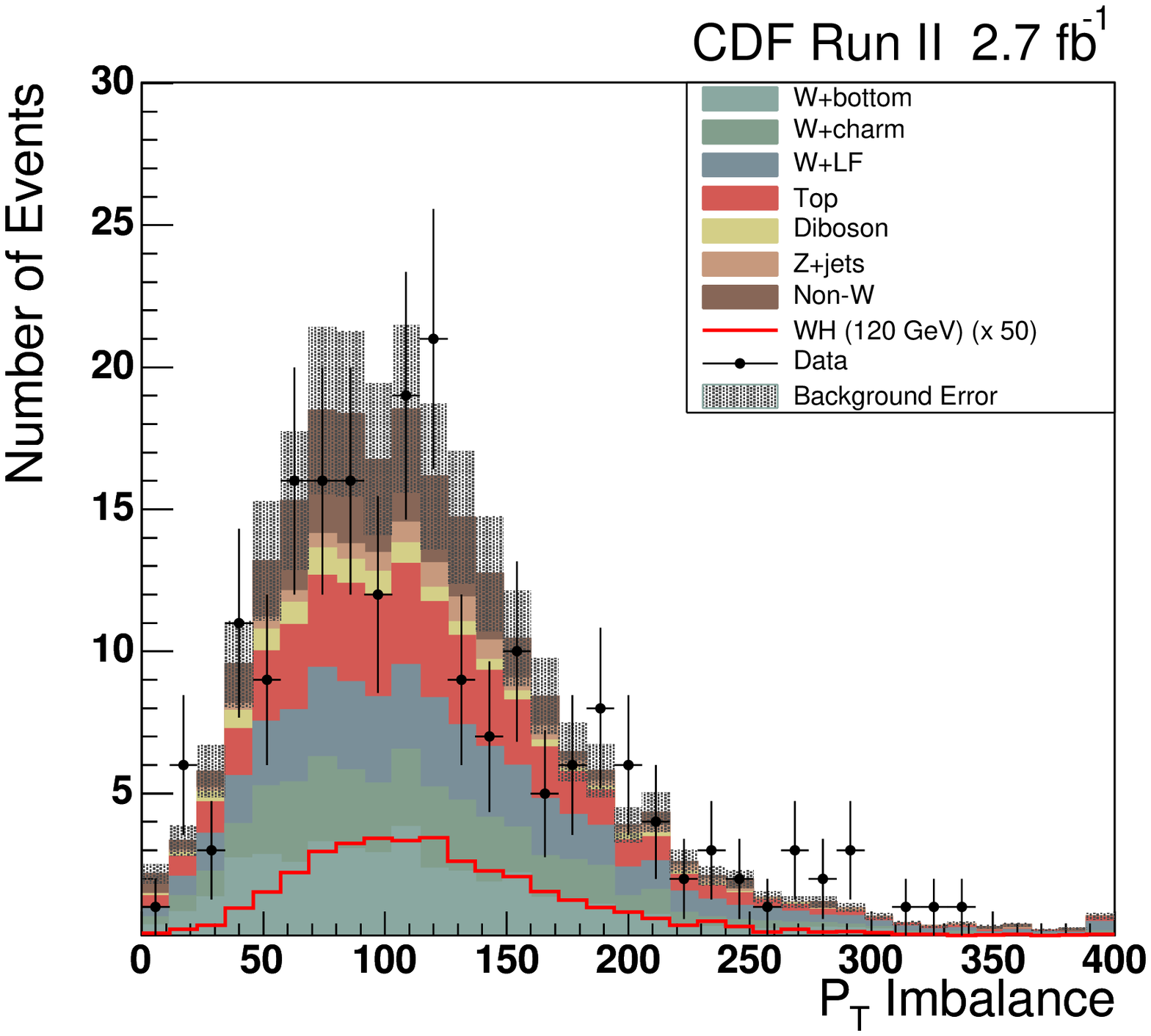}
    \caption{ Neural network input distributions for isolated track
      $W+2$ jet events in the one SECVTX  region. The distributions
      shown are $M_{l\nu j}^{min}$ (left) and $P_T$ Imbalance (right).
      The differences in the shape are consistent with 
      the uncertainty on our QCD model.
      \label{fig:NNinput5}
    }

  \end{center}
\end{figure}

\begin{figure}[htbp]
  \begin{center}

    \includegraphics*[width=0.48\textwidth]{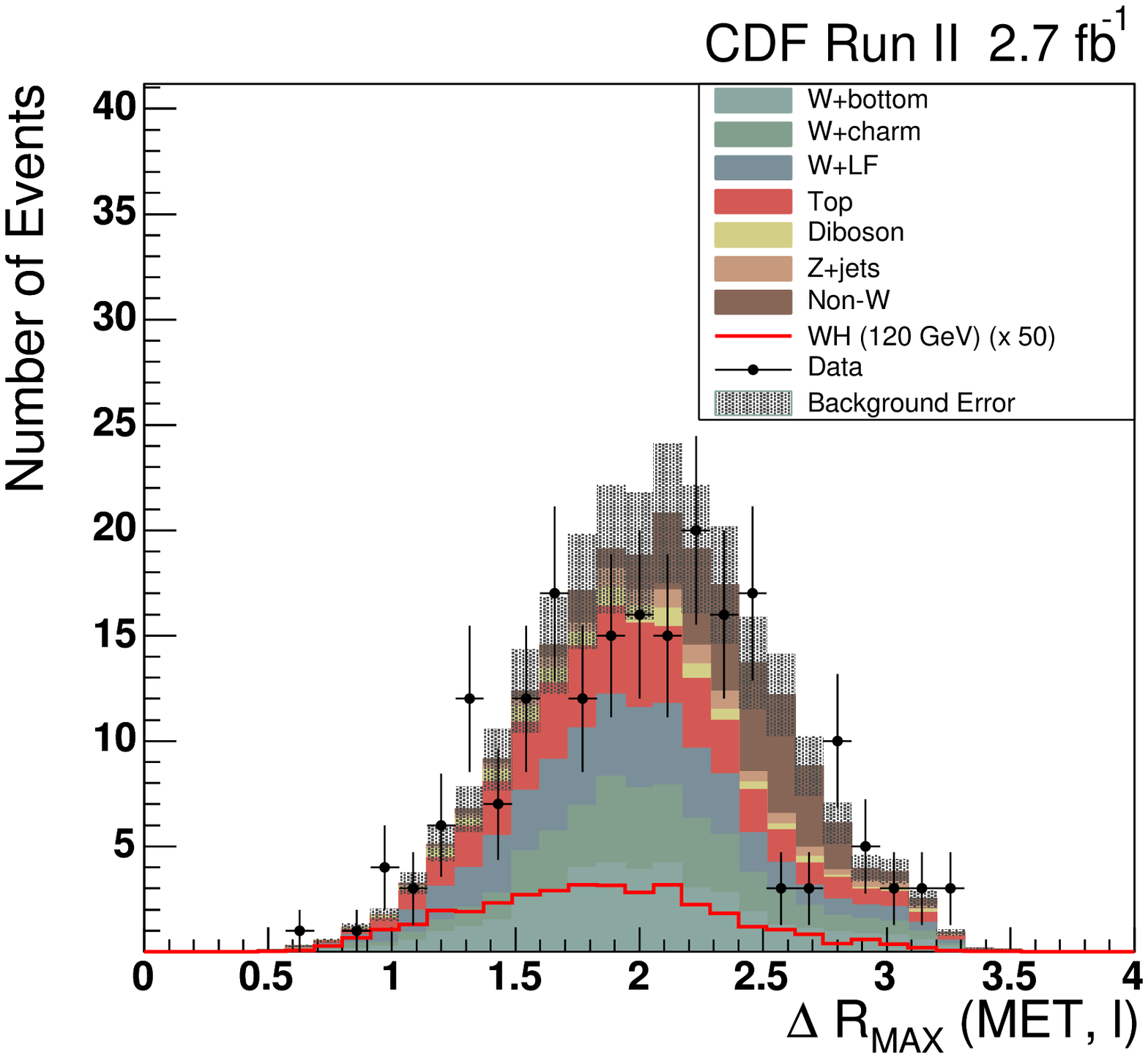} 
    \includegraphics*[width=0.48\textwidth]{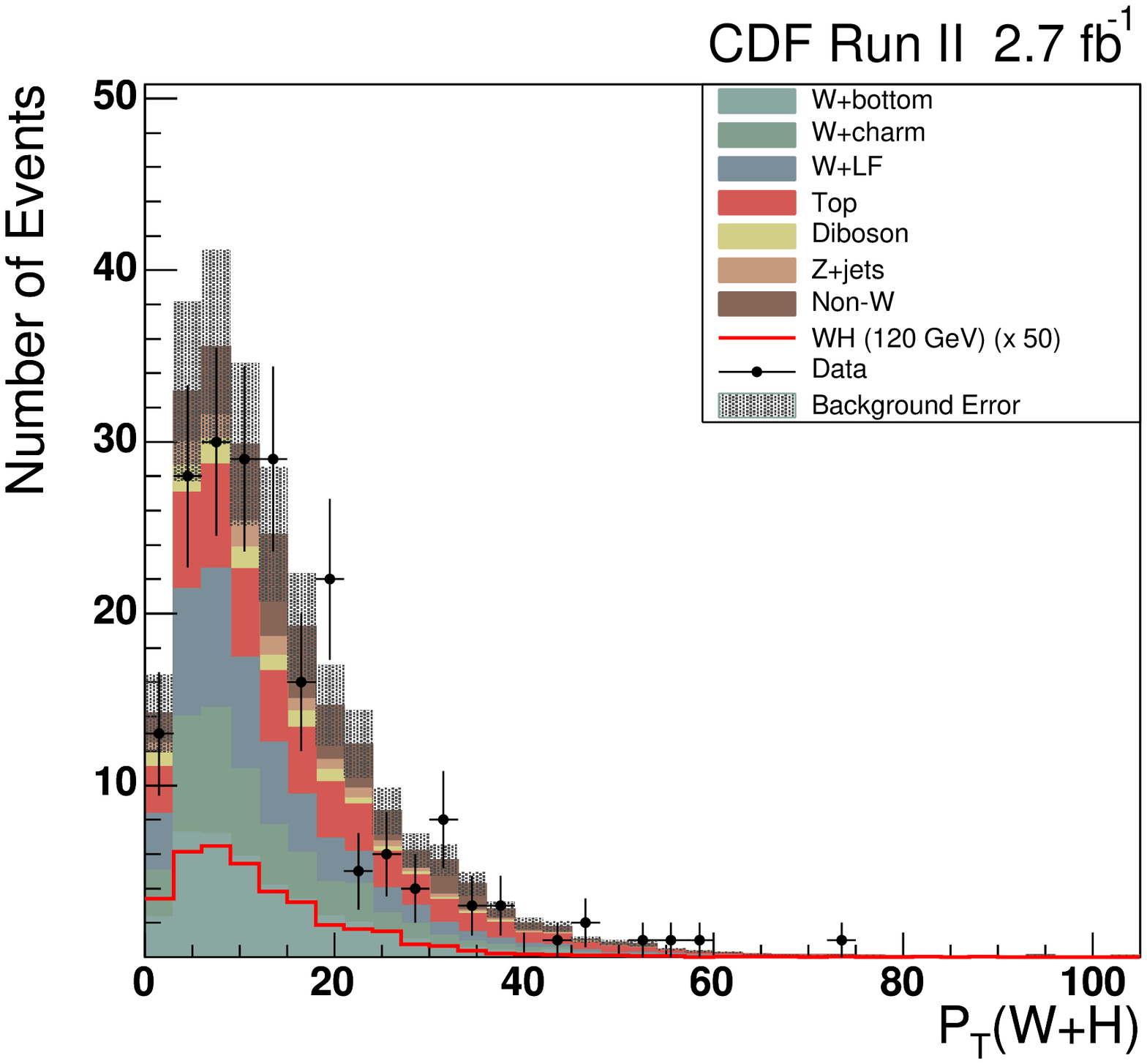}
    \caption{Neural network input distributions for isolated track
      $W+2$ jet events in the one SECVTX region. The distributions
      shown are $\Delta R$(lepton-$\nu_{max}$) (left), $P_T(W+H)$
      (right). The differences in the shape are consistent with 
      the uncertainty on our QCD model.
      \label{fig:NNinput6}
    }

  \end{center}
\end{figure}

Figures~\ref{fig:NNinput1} through \ref{fig:NNinput3} show the six
neural network input variables for isolated track events in the pretag
control region. The plots show that our background model describes the
data reasonably for all the neural network input variables. The
modeling is not ideal in regions that have a large amount of QCD, such
as the region around $\Delta R_{MAX}(MET, l) = 2.5 $ in
Figure~\ref{fig:NNinput3} and the region around $M_{l\nu j}^{min} = 50$ in
Figure~\ref{fig:NNinput2}. Figures~\ref{fig:NNinput4} through \ref{fig:NNinput6}
show that these differences are less significant after removing some
of the QCD contamination with $b$-tagging. The hashed region in
Figures~\ref{fig:NNinput4} through \ref{fig:NNinput6} indicates uncertainty on
the background estimate. Taking into account the uncertainty on the
background estimate, this modeling is reasonable for the isolated
track neural network input variables. 

We studied the impact of QCD shape modeling in the tight lepton
sample. We did not expect the QCD shape to have a large impact on the
sensitivity because the neural network was not trained with QCD
events. We found that the large QCD normalization uncertainty (40\%)
accounted for the small variations that arose from using an
alternative QCD model with different kinematics.  Based on the tight
lepton studies, we assume that the impact of QCD shape modeling on
isolated track sample is also small compared to the QCD normalization
uncertainty. This is not an aggressive assumption since the isolated
track sample only accounts for 20\% of the total sensitivity.

The tight lepton categories also show good agreement with the previous
publication \cite{WH2FB}.

%
%

\section{Limit on Higgs Boson Production Rate}
\label{sec:limit}

\begin{figure}[htbp]
  \begin{center}
    \includegraphics*[width=0.48\textwidth]{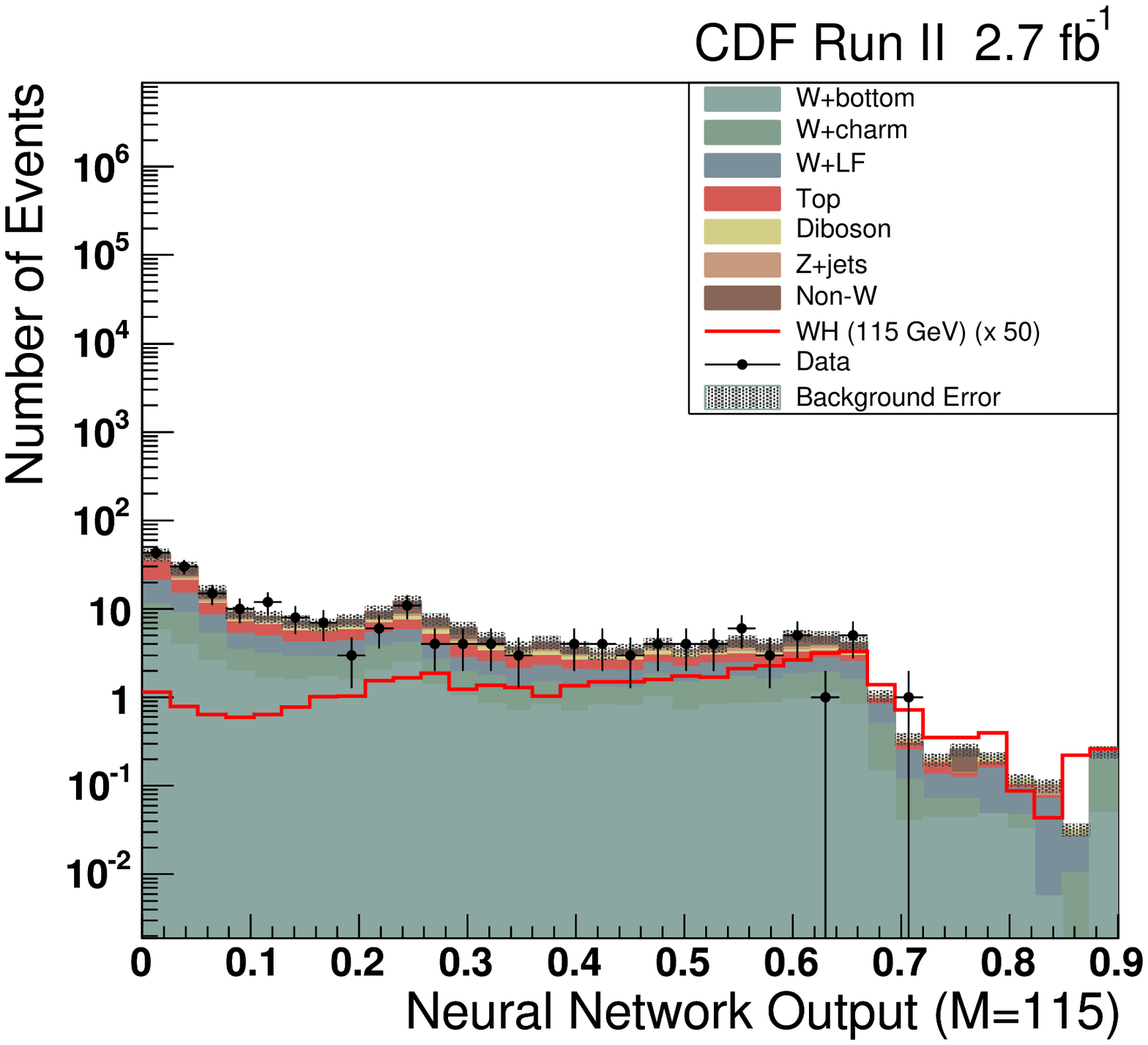}
    \includegraphics*[width=0.48\textwidth]{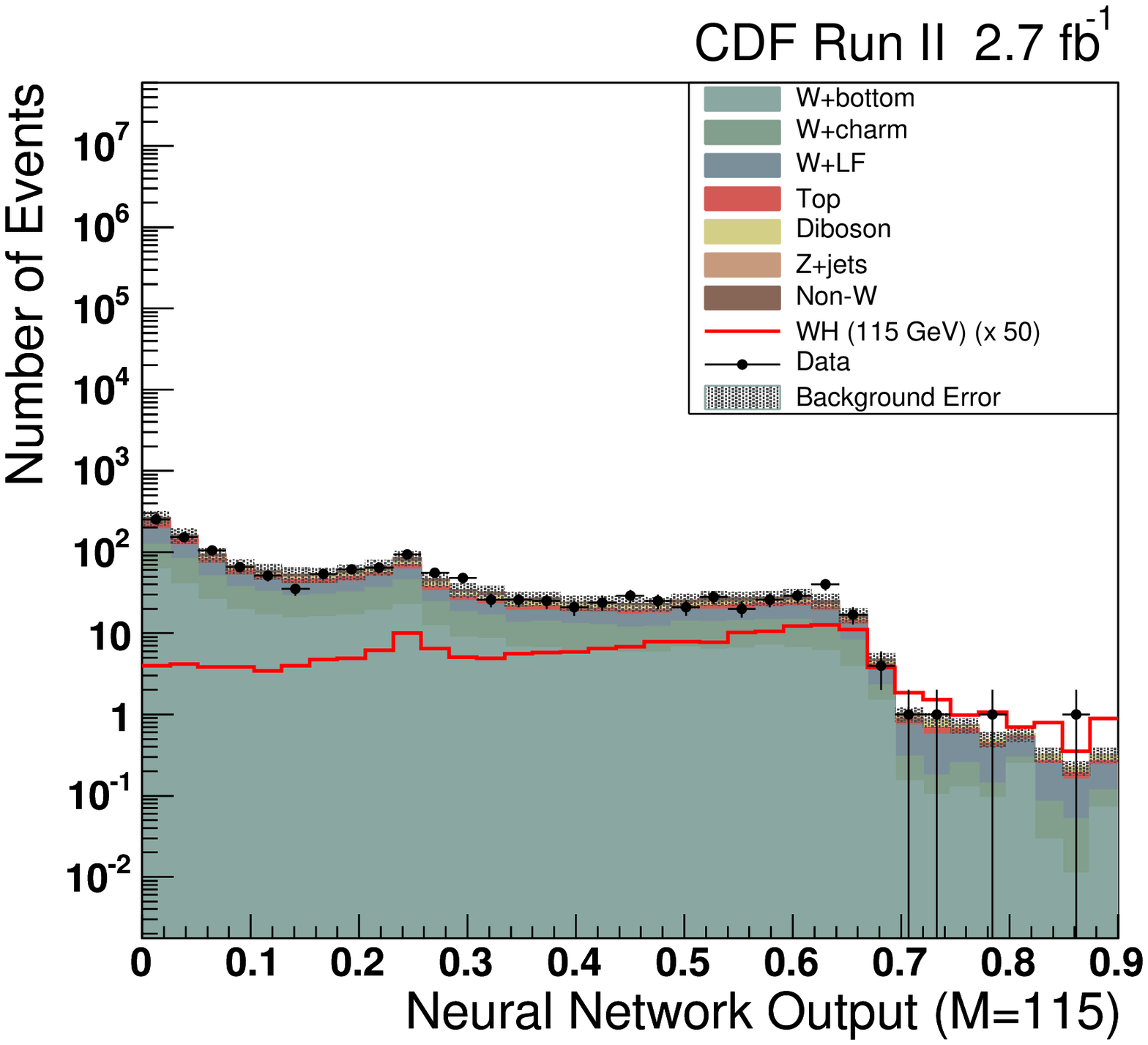}
    
    \caption{Neural Network output distributions for events with one
      Secvtx tag.  The neural network output is close to zero for
      ``background-like'' events, and close to one for ``signal-like''
      events.  The open red curve shows the expected distribution of $WH$
      Monte Carlo events.  The $WH$ expected curve is normalized to 50 times the standard
      model expectation.  The plots show isolated track events (left)
      and lepton triggered events (right).  }
    \label{fig:nnOutput_eq1tag}
  \end{center}
\end{figure}

\begin{figure}[htbp]
  \begin{center}
    \includegraphics*[width=0.48\textwidth]{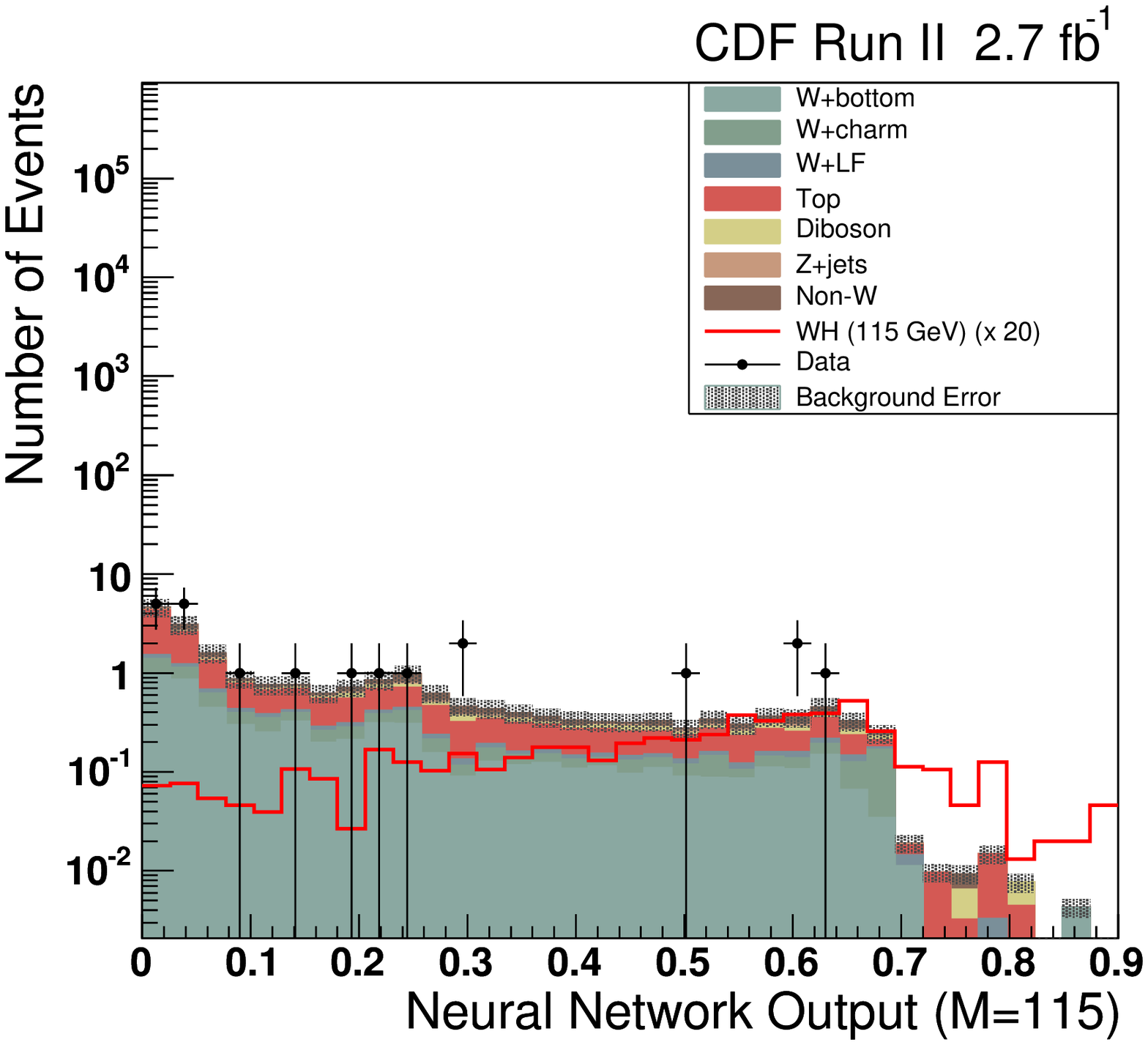}
    \includegraphics*[width=0.48\textwidth]{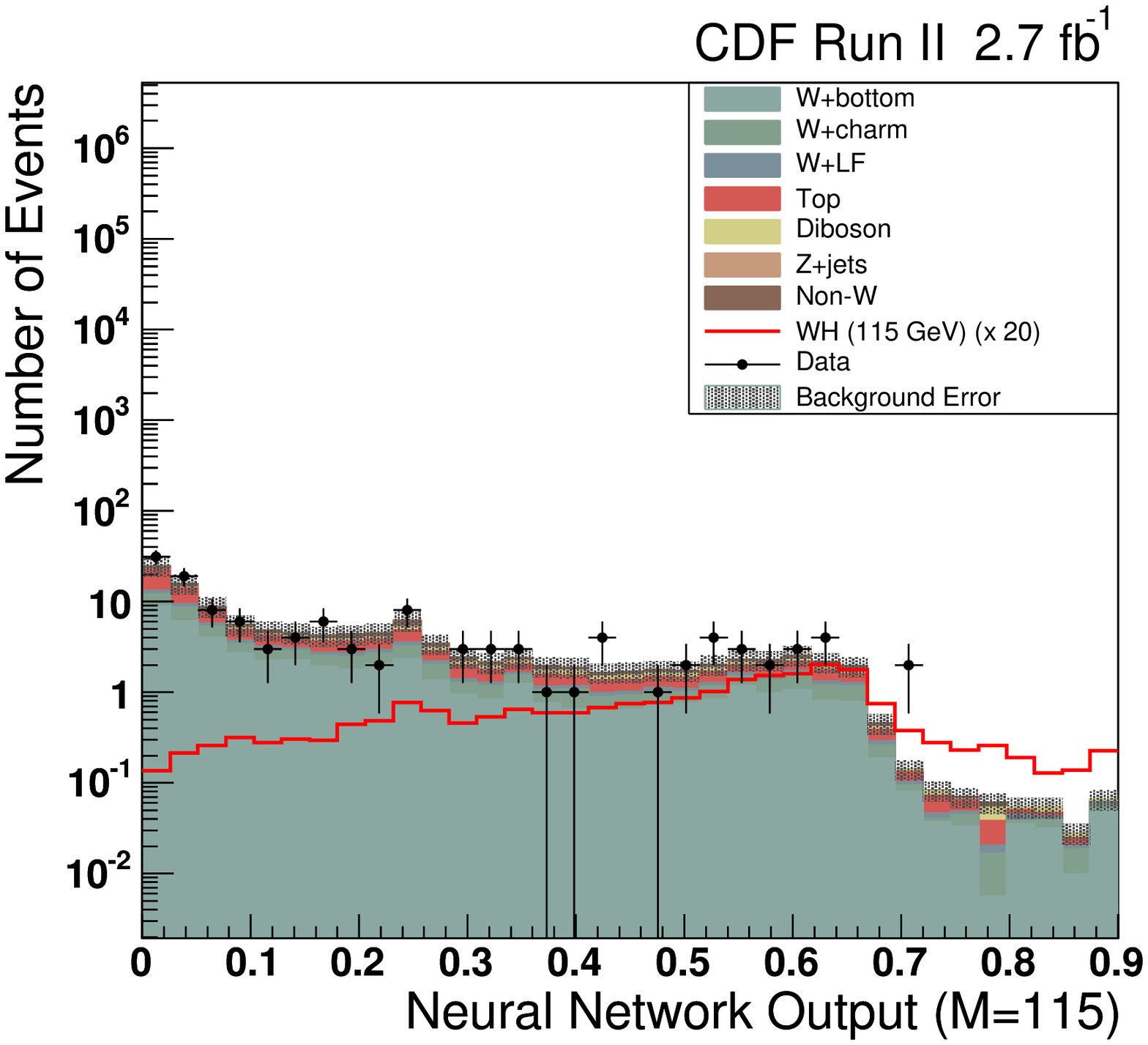}
    
    \caption{Neural Network output distributions for events with one
      secvtx tag and one jet probability tag.  The neural network
      output is close to zero for ``background-like'' events, and
      close to one for ``signal-like'' events.  The open red curve
      shows the expected distribution of $WH$ Monte Carlo events.  The
      $WH$ expected curve is normalized to 50 times the standard model
      expectation.  The plots show isolated track events (left) and
      lepton triggered events (right).  }
    \label{fig:nnOutput_stagJP}
  \end{center}
\end{figure}

\begin{figure}[htbp]
  \begin{center}
    \includegraphics*[width=0.48\textwidth]{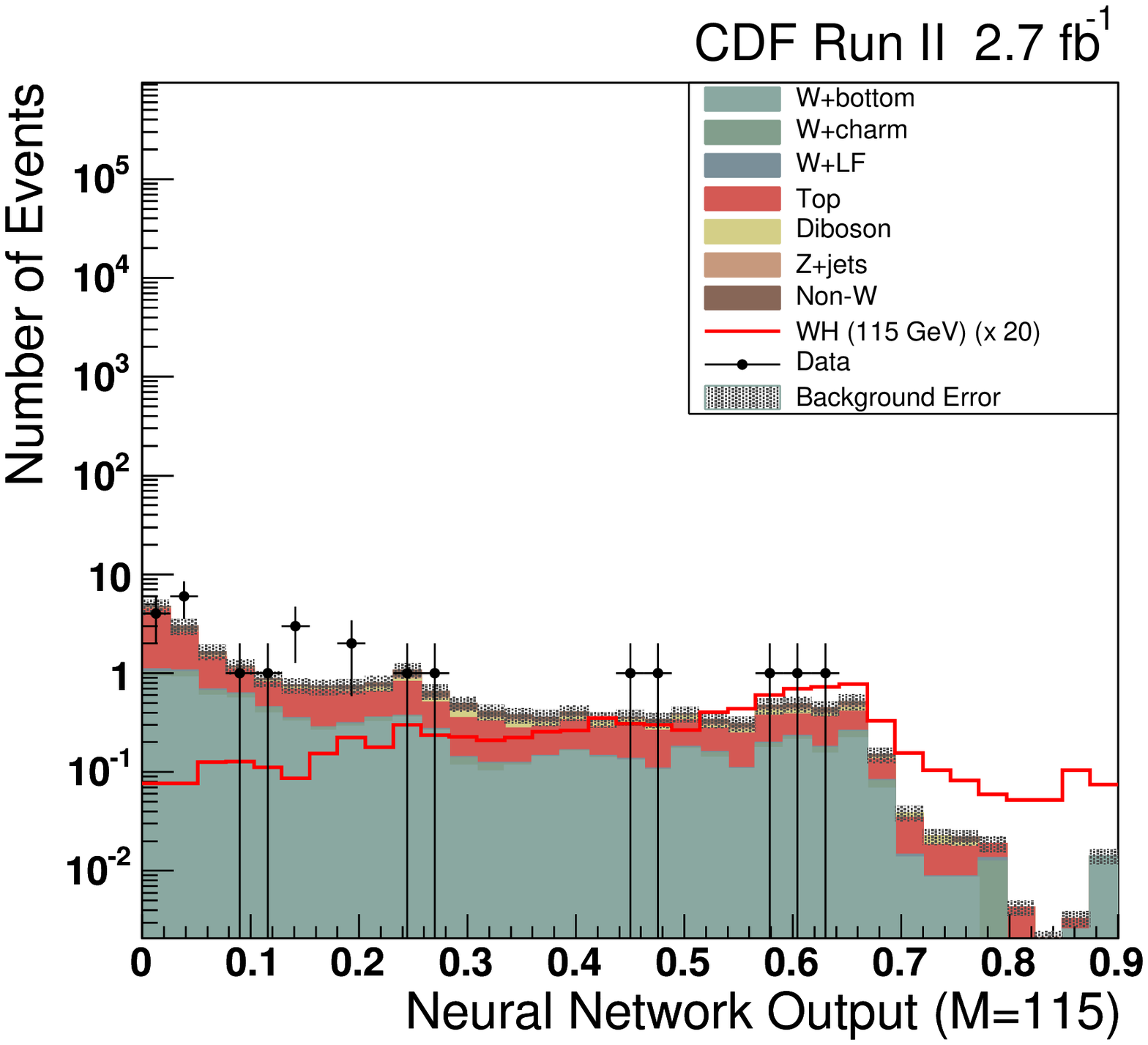}
    \includegraphics*[width=0.48\textwidth]{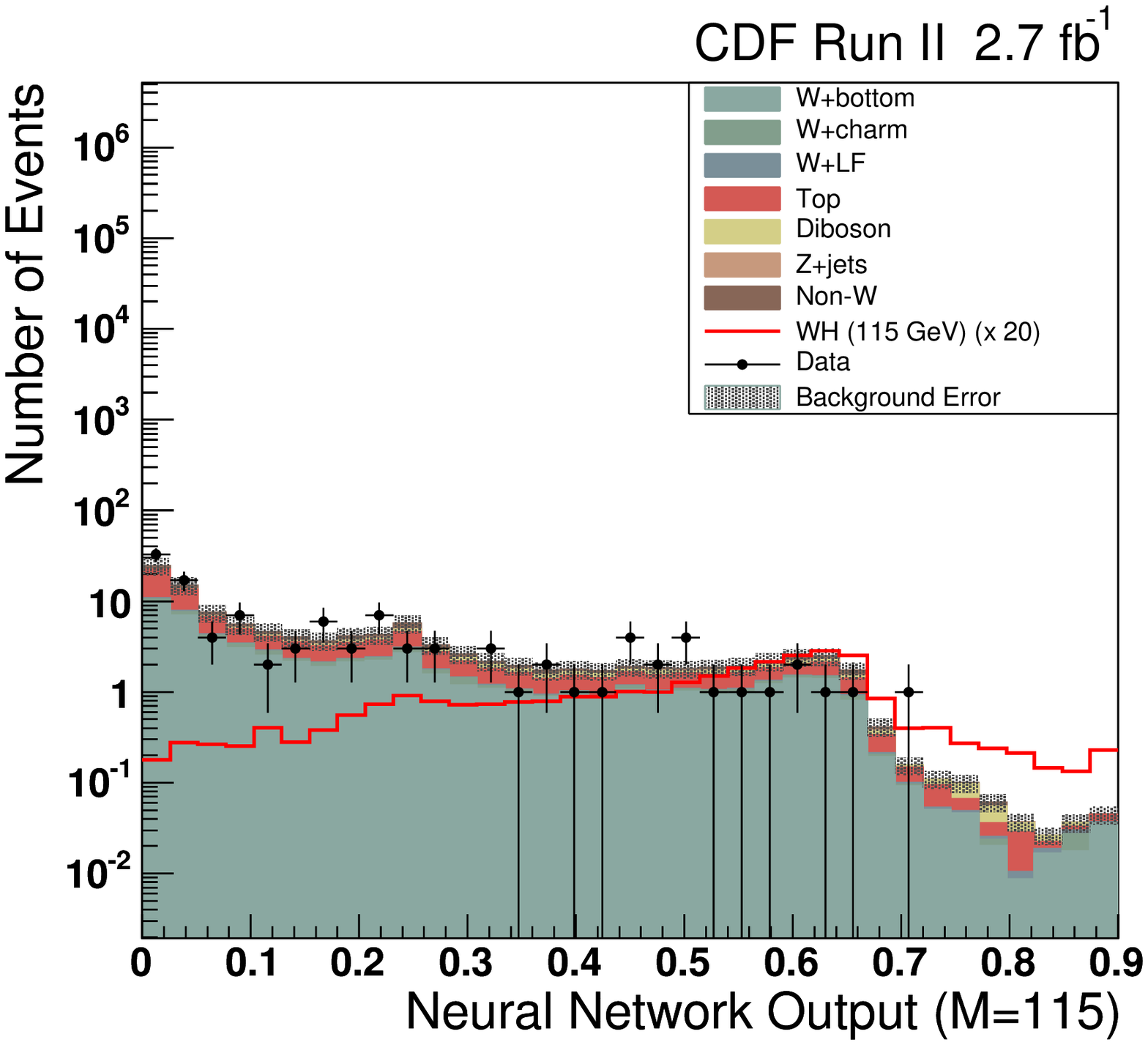}
    
    \caption{Neural Network output distributions for events with two
      secvtx tags.  The neural network output is close to zero for
      ``background-like'' events, and close to one for ``signal-like''
      events.  The open red curve shows the distribution of $WH$
      events.  The $WH$ curve is normalized to 50 times the standard
      model expectation.  The plots show isolated track events (left)
      and lepton triggered events (right).  }
    \label{fig:nnOutput_dtag}
  \end{center}
\end{figure}

\begin{figure}
  \begin{center}
    \includegraphics[width=0.9\textwidth]{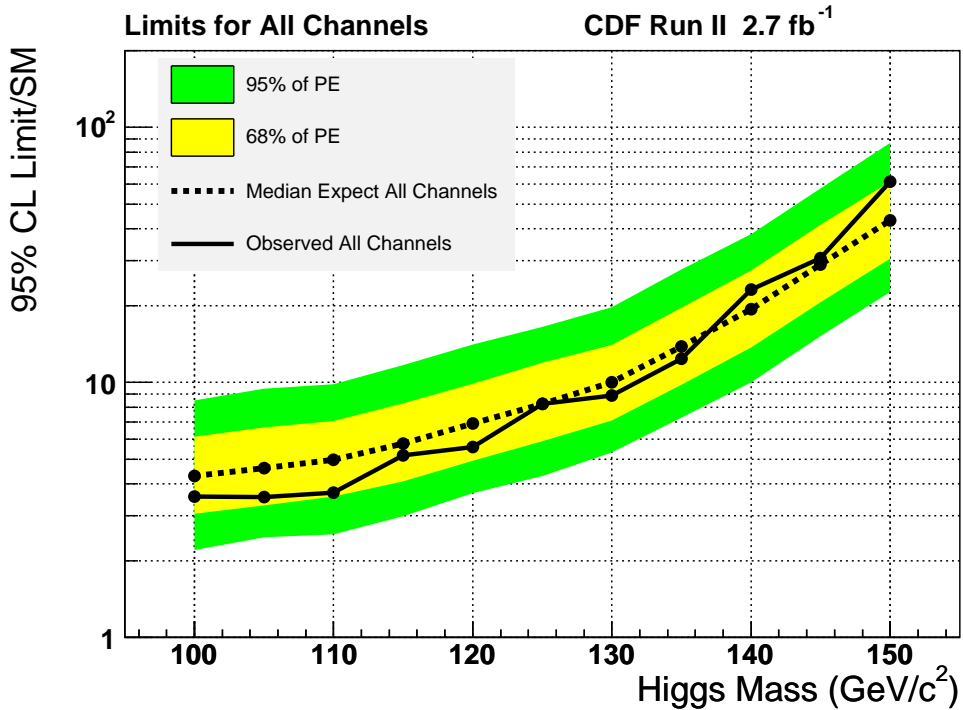}
   \caption{95\% confidence level upper limit on
     $\sigma(p\bar{p}\rightarrow WH)\cdot {\cal B}(H\rightarrow
     b\bar{b})$, expressed as a ratio to the standard model
     expectation.  The limits were obtained using an integrated
     luminosity of $2.7$~fb$^ {-1}$ and analyzing both lepton triggered
     and \MET + 2 jet triggered events.  The dashed line indicates the
   median expected limit. The yellow and green  regions encompass
   the limits in  68\% and 95\% of pseudo-experiments, respectively.
   The solid line shows the observed limits.}
   \label{fig:upperLimit}
  \end{center}
 \end{figure}

We search for an excess of Higgs signal events in our neural network
output distributions using a binned likelihood technique. Figures
~\ref{fig:nnOutput_eq1tag} through~\ref{fig:nnOutput_dtag} show the
neural network output distributions for events in different lepton and
tag categories.  We use the same likelihood expression and
maximization technique as the prior CDF result \cite{WH2FB} and
described in \cite{MCLIMIT}. We maximize the likelihood, fitting for a
combination of Higgs signal plus backgrounds. We find no evidence for
a Higgs boson signal in our sample, and so we set 95\% confidence level upper
limits on the $WH$ cross section times branching ratio:
$\sigma(p\bar{p}\rightarrow W^{\pm}H)\cdot {\cal B}(H\rightarrow
b\bar{b})$.  We compare our observed limits to our expected
sensitivity by creating pseudo-experiments with pseudo-data
constructed from a sum of background templates.  Our expected and
observed limits are shown in Fig.~\ref{fig:upperLimit} and
Table~\ref{table:LimitCombined}. The limits are expressed as a
function of the Higgs boson mass hypothesis.

The likelihood technique accommodates the uncertainties on our
background estimate by letting the overall background prediction float
within Gaussian constraints. We use a different set of background and
signal neural network template shapes for each combination of lepton
type and tag category as a separate channel in the likelihood. We
correlate the systematic uncertainties appropriately across different
lepton types and tag categories.


\begin{table}
\begin{center}
  \begin{tabular}{ccc}
    \hline
    \hline  
    \multicolumn{3}{c}{CDF Run II Preliminary 2.7 fb$^{-1}$}\\
    \multicolumn{3}{c}{Limits for Combined Lepton and Tag Categories}\\
    \multicolumn{3}{c}{in units of SM cross sections}\\
    \hline 
    M(H)  & Observed Limit (x SM) & Expected Limit (x SM) \\
    \hline

    100 & 3.6  & 4.3 \\
    105 & 3.6  & 4.6 \\
    110 & 3.7  & 5.0 \\
    115 & 5.2  & 5.8 \\
    120 & 5.6  & 6.9 \\
    125 & 8.2  & 8.2 \\
    130 & 8.9  & 10.0 \\
    135 & 12.4 & 13.8 \\
    140 & 23.1 & 19.4 \\
    145 & 30.6 & 28.9 \\
    150 & 61.1 & 43.2 \\
    
    \hline
    \hline
  \end{tabular}
  \caption{Expected and observed limits as a function of Higgs mass
    for the combined search of Tight Lepton and Isotrk events, including
    all tag categories. The limits are expressed in units of Standard Model
    $WH$ cross sections. }
  \label{table:LimitCombined}
\end{center}
\end{table}

\section{Conclusions}
\label{sec:conclusions}

Our limit on $WH$ production improves on the previous result by using
more integrated luminosity and extending the lepton identification
with isolated tracks. 
The increase in luminosity from 1.9 fb$^{-1}$ to 2.7 fb$^{-1}$
increases the sensitivity by $\sim$20\%. Using isolated track events
provides a $\sim$25\% increase in acceptance above the prior
analysis. The new isolated track events combined with minor
improvements in background rejection yield a overall $\sim$15\%
increase in estimated sensitivity. Our expected limits are expressed
as a ratio to the standard model production rate. The expected limits
vary from 4.3 to 43.2 for Higgs masses from 100 to 150 GeV/$c^{2}$,
respectively. We find no evidence for Higgs production in the data,
and set observed limits at 3.6 to 61.1 for Higgs masses from 100 to
150 GeV/$c^{2}$, respectively.

\begin{acknowledgments}

  We thank the Fermilab staff and the technical staffs of the
  participating institutions for their vital contributions. This work
  was supported by the U.S. Department of Energy and National Science
  Foundation; the Italian Istituto Nazionale di Fisica Nucleare; the
  Ministry of Education, Culture, Sports, Science and Technology of
  Japan; the Natural Sciences and Engineering Research Council of
  Canada; the National Science Council of the Republic of China; the
  Swiss National Science Foundation; the A.P. Sloan Foundation; the
  Bundesministerium f\"ur Bildung und Forschung, Germany; the Korean
  World Class University Program, the National Research Foundation of
  Korea; the Science and Technology Facilities Council and the Royal
  Society, UK; the Institut National de Physique Nucleaire et Physique
  des Particules/CNRS; the Russian Foundation for Basic Research; the
  Ministerio de Ciencia e Innovaci\'{o}n, and Programa
  Consolider-Ingenio 2010, Spain; the Slovak R\&D Agency; the Academy
  of Finland; and the Australian Research Council (ARC).

\end{acknowledgments}

\bibliography{reference}

\end{document}